\documentclass[aps, prb, groupaddress, reprint, showpacs]{revtex4-1}

\usepackage{amsfonts}
\usepackage{amsmath}
\usepackage{amssymb, bm, mathrsfs}
\usepackage{graphicx, epstopdf, color}
\usepackage{soul}
\usepackage{appendix}
\usepackage{hyperref}

\begin{document}

\title{Generalized Fowler-Nordheim field-induced vertical electron emission model for two-dimensional materials}

\author{Yee Sin Ang}
\email{yeesin\_ang@sutd.edu.sg}
\affiliation{SUTD-MIT International Design Center, Singapore University of Technology and Design, 8 Somapah Road, Singapore 487372}

\author{M. Zubair}
\affiliation{SUTD-MIT International Design Center, Singapore University of Technology and Design, 8 Somapah Road, Singapore 487372}

\author{K. J. A. Ooi}
\affiliation{SUTD-MIT International Design Center, Singapore University of Technology and Design, 8 Somapah Road, Singapore 487372}

\author{L. K. Ang}
\email{ricky\_ang@sutd.edu.sg}
\affiliation{SUTD-MIT International Design Center, Singapore University of Technology and Design, 8 Somapah Road, Singapore 487372}

\begin{abstract}
	 
Current theoretical description of field-induced electron emission remains mostly bounded by the classic Fowler-Nordheim (FN) framework developed nearly one century ago. For the emerging class of two-dimensional (2D) materials, many basic assumptions of FN model become invalid due to their reduced dimensionality and exotic electronic properties. In this work, we develop analytical and semi-analytical models of field-induced vertical electron emission from the surface of 2D materials by explicitly taking into account the reduced dimensionality, non-parabolic energy spectrum, non-conserving in-plane electron momentum, finite-temperature and space-charge-limited effects. We show that the traditional FN law is no longer valid for 2D materials. The modified vertical field emission model developed here provides better agreement with experimental results. Intriguingly, a new high-field regime of \emph{saturated surface field emission} emerges due to the reduced dimensionality of 2D materials. A remarkable consequence of this saturated field emission effect is the absence of space-charge-limited current normally expected at high field in three-dimensional bulk material.

\end{abstract}

\maketitle

\section{Introduction}

The formulation of electron emission theory based on Sommerfeld's free electron model of metals \cite{sommerfeld} represent one of the earliest successes of quantum mechanics in solid state physics \cite{hoddeson}. The \emph{thermionic emission} (TE) of hot electrons above the surface potential barrier and the sub-barrier quantum mechanical tunnelling of cold electron via \emph{field emission} (FE) are governed, respectively, by the Richardson-Dushman (RD) \cite{RD} and the Fowler-Nordheim (FN) laws \cite{FN}, 
\begin{subequations}\label{T_F_E}
	\begin{equation}\label{J_TE}
	\mathscr{J}_{\text{RD}} = A_{\text{RD}} T^2 \exp \left[ -\frac{\Phi_B }{k_BT} \right],
	\end{equation}
	\begin{equation}\label{J_FE}
	\mathcal{J}_{\text{FN}} = A_{\text{FN} } F^2 \exp \left[ - B_{\text{FN}}\frac{ \Phi_B ^{3/2}}{F} \right],
	\end{equation}
\end{subequations}
where $\Phi_B$ , $T$, $F$, $A_{\text{RD}}$, $A_{\text{FN}}$ and $B_{\text{FN}}$ are the work function, temperature, electric field strength, Richardson constant, the first and the second FN constants, respectively \cite{emission_const}. 

The RD and the FN laws follow a remarkable symmetry -- both are composed of a material-dependent pre-factor multiplied by a universal exponential factor of $\exp ( -\Phi_B/k_BT )$ for TE and $\exp ( - B_{FN}\Phi_B^{3/2}/k_BT )$ for FE. 
The exponential terms provide tell-tale signatures of the underlying emission mechanism since the $\mathscr{J}$-$T$ and the $\mathcal{J}$-$F$ characteristics are almost exclusively determined by temperature and electric field. 
Thus, Eq. (\ref{T_F_E}) is extraordinarily versatile and can be nearly universally applied to analyse the electron emission characteristics for a wide range of materials by using the \emph{Arrhenius plot}: $\ln (\mathscr{J}/T^2)$ versus $1/T$ for TE or the \emph{FN plot}: $\ln(\mathcal{J}/F^2)$ versus $1/F$ for FE. 
Nonetheless, much of the material-dependent physics contained in the pre-factor remains obscured in the experimental data when analysed using Eq. (\ref{T_F_E}). This poses a potential danger of misinterpreting the emission physics in unconventional materials especially when their physical properties do not permit the use of Eq. (\ref{T_F_E}).

Although the classic RD and FN models for bulk materials have been continually refined over past decades \cite{p_zhang, TF, MG, burgess, stratton, forbes, forbes2, forbes3, forbes4, jensen, jensen2, jensen3, kyritsakis, sd_liang, wei, wei2, edgcombe, holgate}, their validity for the emerging class of two-dimensional (2D) materials have been questioned recently \cite{liang, sinha, ang, IEDM2016, misra, misra2, liang2}.
Based on a continuum emission picture, Liang and Ang have shown that graphene thermionic emitter exhibits unconventional $T^3$ scaling in the pre-factor \cite{liang, liang2}. 
The Liang-Ang continuum model is generalized to the case of narrow-gap semiconductor and few-layer graphene by Ang and Ang \cite{ang}.
To model the charge transport across a graphene-based Schottky contact, Sinha and Lee adopted an alternative `time'-constant-based TE model \cite{sinha}, which contains a polynomial pre-factor of $T^2(1 + \Phi_B/k_BT)$ and an empirical `time'-like parameter, $\tau$, in the emission current density. 
Albeit the very different underlying physical pictures, both Liang-Ang and Sinha-Lee TE models agree reasonably well with experiments \cite{sinha, IEDM2016, massicotte}.

\begin{figure*}
	\includegraphics[scale=0.55]{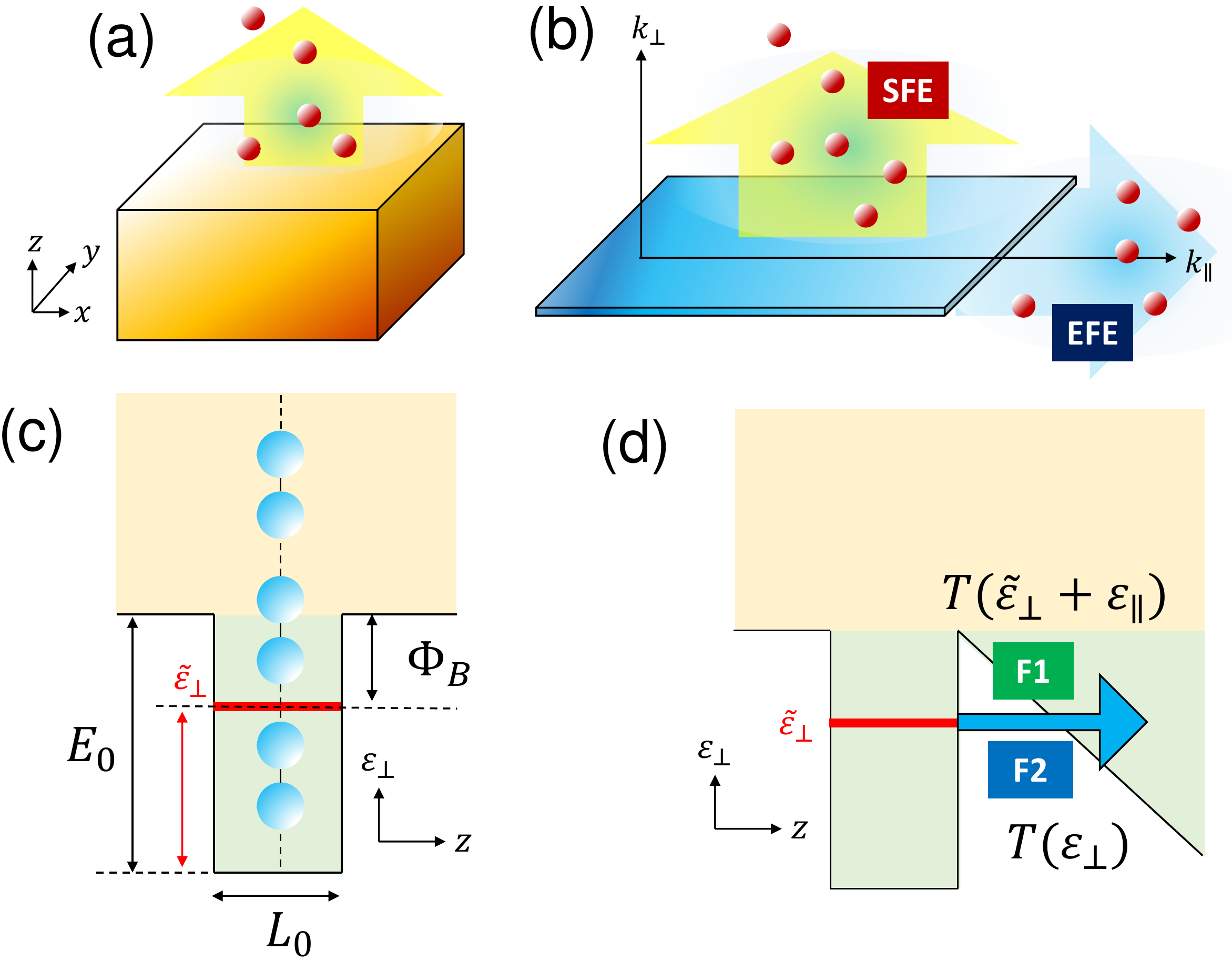}
	\caption{Electron emission from (a) three-dimensional bulk; and (b) two-dimensional plane. For emission from 2D plane, two emission configurations are possible, i.e. edge field emission (EFE) and surface field emission (SFE). (c) Model of electrons in 2D material as a quantum square well. The terms $\tilde{\varepsilon}_{\perp}$, $L_0$, $\Phi_B$ and $E_0$ denotes the bound state energy measured with respect to well bottom, quantum well width, surface potential barrier and quantum well depth, respectively. (d) Two field emission path ways in 2D materials: (F1) non-conserving lateral momentum (NCLM) field emission and (F2) conserving lateral momentum (CLM) field emission. }
\end{figure*}

Despite significant progress in the theoretical development of TE, cold electron FE theory remains relatively less-explored \cite{MRS}. 
For traditional bulk material, FE does not sensitively rely on the direction of emission for materials with isotropic energy dispersion [Fig. 1(a)]. 
In contrast, the thin-film nature of 2D materials breaks the symmetry of emission configuration and leads to two distinct subtypes: (i) \emph{edge field emission} (EFE) where electrons are emitted along the 2D plane; and (ii) \emph{surface field emission} (SFE) where electrons are emitted normally from the 2D-plane [see Fig. 1(b)].  
In EFE \cite{wang_edge, kumar, moussa, xu, wu_GVT, sun, song, carbon}, the electrons has a continuous in-plane energy dispersion component along the emission direction while in SFE, the electron dynamics along the emission direction is quantized due to the quantum confinement by surface potential \cite{rz_wang, wei3}.
Nonetheless, the traditional FN law is employed in the vast majority of graphene FE experiments \cite{malesevic, eda, qian, wu, huang, yamaguchi, santandrea, wang_edge, kumar, palnitka, soin, kleshch, moussa, xu, wu_SFE, yusop, bartolomeo, Iemmo} albeit the fact that the underlying FE mechanism in 2D material differs significantly from the bulk counterpart. 
Modified EFE theory based on a 2D emitter model is proposed \cite{qin} and verified experimentally \cite{xiao} to exhibit an unconventional $F^{3/2}$-scaling in the prefactor.
Such EFE model is further refined by Wang et al to include the geometrical edge effect and the relativistic energy dispersion of graphene \cite{wang}. 

Thus far, the physical model underlying SFE from 2D materials remains incomplete. In this work, we develop analytical and semi-analytical models of field-induced vertical electron emission from the surface of 2D materials by explicitly taking into account their reduced dimensionality, non-parabolic energy spectrum, non-conservation of lateral momentum (i.e. momentum component lying in the plane of 2D material), finite-temperature and space-charge-limited effects. 
Our model exhibits a significant departure from the traditional FN law, thus signaling the breaks down of FN paradigm in 2D materials.
From our model, an unconventional $\mathcal{J}$-$F$ scaling relation is found, i.e.
\begin{equation}
\mathcal{J}_{\text{2D}} \propto \exp\left( -B_{FN}\frac{ \Phi_B^{3/2} }{F} \right).
\end{equation}
Here, the pre-factor is $F$-independent, rather than the $F^2$-dependence in the traditional FN law [see Eq. (\ref{J_FE})]. Correspondingly, the FN-plot is modified as: $\ln \left(\mathcal{J}\right)$ versus $1/F$, which is shown to provide a better agreement with experimental data (see Fig. 4 below).
Intriguingly, the reduced dimensionality of 2D materials bestows a new regime of \emph{saturated surface field emission} at high electric field. 

For FE from 3D bulk materials, when its current density is sufficiently large, the built-up of space charge in the vacuum gap can lead to a transition from field-induced quantum mechanical tunneling to space-charge-limited (SCL) flow \cite{barbour, lau, kelvin2, forbes_trans, p_zhang2}, where the latter is governed by the Child-Langmuir (CL) law \cite{CL}.
The solid-state counter part of SCL effect in 2D Dirac semiconductor has been recently reported to exhibit an unconventional current-voltage scaling relation \cite{ang_DiracSCLC}. 
For 2D material field emitter, the following questions arise: can the emission current from 2D materials evolve from field emission to semiclassical SCL flow? 
Thus far, this question has not been studied both theoretically or experimentally yet. 
Here we address this question by constructing a general emission model -- covering both FE and SCL regimes -- through the combination of 2D material FE models and 1D Poisson equation \cite{lau}.
We show that the dimensionality-induced saturated emission results in a remarkable consequence: the space-charge-limited flow of electrons can be fully eliminated in 2D-material-based surface field emitter at nanometer and micrometer scales with practical range of applied voltage. 

The saturated FE and the peculiar absence of space-charge effect shall have profound impacts on the characterization and design of 2D-material-based field emitter and vacuum nanoelectronics. 
Furthermore, as field-induced vertical electron emission from the 2D planar surface represents one of the central transport mechanisms in solid-state 2D-material-based electrical contact \cite{allain, xu_rev, bartolomeo2, jariwala}, the models developed here shall provide a theoretical basis in the analysis of charge transport phenomena in such interfaces.
%

\section{Overview of electron emission formalism: From bulk to 2D materials}

\subsection{Electron emission from bulk material}

The general form of the electrical current density can be written as
\begin{equation} \label{J}
	J = \frac{1}{\mathcal{V}} \sum_{\mathbf{k}} \mathbf{v}(\mathbf{k})  \mathcal{T} (\mathbf{k})  f(\mathbf{k}),
\end{equation} 
where $\mathbf{k}$, $\mathbf{v}(\mathbf{k})$, $\mathcal{T}(\mathbf{k})$, $f(\mathbf{k})$ and $\mathcal{V}$ are wavevector, carrier group velocity, transmission probabilities, carrier distribution function and total volume of the emission system, respectively. We partition the emitted electron energy and wavevector as followed:
\begin{eqnarray} \label{part}
\varepsilon\left(\mathbf{k}\right) &=& \varepsilon_\parallel\left(\mathbf{k}_\parallel\right) + \varepsilon_\perp \left(\mathbf{k}_\perp\right)  \nonumber \\
\mathbf{k} &=& \mathbf{k}_\parallel + \mathbf{k}_\perp,
\end{eqnarray}
where $\varepsilon_\parallel$ ($\varepsilon_\perp$) is the energy component transversal to (along) the emission direction, and $\mathbf{k}_\parallel$ ($\mathbf{k}_\perp$) is the wavevector component transversal to (along) the emission direction [see Fig. 1(b)]. For a three-dimensional (3D) bulk material, the emission current density can be written as
\begin{eqnarray}\label{gen_J}
J_{\text{bulk}} &=& \frac{eg_{s,v}}{(2\pi)^3 }\int d \mathbf{k}_\parallel \int_0^\infty dk_\perp v(k_\perp) \mathcal{T}(k_\perp) f(\mathbf{k}) \nonumber \\
&=& \frac{e}{2\pi \hbar}\int_0^{\infty} D(\varepsilon_\parallel)d \varepsilon_\parallel \int_{0}^{\infty} \mathcal{T}(\varepsilon_\perp) f(\varepsilon) d\varepsilon_\perp,
\end{eqnarray}
where the \emph{density of states} (DOS) in the plane lateral to the emission direction is defined as $D(\varepsilon_\parallel) = g_{s,v} k_{\parallel} dk_\parallel / 2\pi$ for an isotropic lateral energy dispersion and $g_{s,v}$ is the spin-valley degeneracy.
In obtaining Eq. (\ref{gen_J}), we have taken the continuum limit of $\mathcal{V}^{-1}\sum_{\mathbf{k}} \to (2\pi)^{-1} \int d^3 \mathbf{k}$ and re-written the group velocity along the emission direction as $v(k_\perp) = \hbar^{-1} d \varepsilon_\perp / dk_\perp$. For TE, the transmission probabilities are set to unity for $\varepsilon_\perp > \Phi_B$ and the carrier distribution function is approximated by semiclassical distribution, 
	\begin{eqnarray}\label{TE_cond}
	\mathcal{T}(\varepsilon_\perp) &=& \Theta(\varepsilon_\perp - \Phi_B), \nonumber \\ f(\varepsilon) &=& \exp\left( -\frac{\varepsilon - \varepsilon_F}{k_BT} \right).
	\end{eqnarray}
For FE, the zero-temperature Fermi-Dirac distribution is employed and the transmission probabilities is calculated via the Wentzel-Kramers-Brillouin (WKB) approximation for triangular barrier, i.e.
	\begin{eqnarray}\label{FE_cond}
	\mathcal{T}(\varepsilon_\perp) &=& D_F \exp\left( \frac{\varepsilon_\perp -\varepsilon_F}{d_F} \right), \nonumber \\
	f(\varepsilon) &=& \Theta\left( \varepsilon_F - \varepsilon \right),
	\end{eqnarray}
where $D_F \equiv \exp \left( - 2\Phi_B/3d_F \right)$, $d_F \equiv e \hbar F/ 2\sqrt{2m\Phi_B}$, $m$ is the electron mass in free-space, $F$ is the electric field strength and $B_{FN} = 4\sqrt{2m}/3e\hbar$ \cite{emission_const}. By combining Eqs. (\ref{gen_J}--\ref{FE_cond}), we obtain the traditional RD and FN relations in Eq. (\ref{T_F_E}).
Note that in taking the transmission probability in the form of $\mathcal{T}(\varepsilon_\perp)$, it is implied that the lateral energy component, $\varepsilon_\parallel$, does not contribute to the $\perp$-directional tunneling, i.e. the lateral momentum component, $k_\parallel$, is strictly conserved.
Thus, the classic RD and FN laws belong to the class of \emph{conversing lateral momentum} (CLM) emission theories. 
As we shall see below, CLM can be relaxed in 2D materials and this leads to distinctive electron emission characteristics.

\subsection{Electron emission from the surface of 2D material}

The traditional FN law [see Eq. (\ref{J_FE}) above] is derived based on the following assumptions: (i) in order to calculate the emission current density, both $\varepsilon_\parallel$ and $\varepsilon_\perp$ are required to be continuously integrable; (ii) the $\varepsilon_\parallel$ is assumed to be parabolic in $\mathbf{k}_\parallel$; (iii) due to CLM, the transmission probability is a function of $\varepsilon_\perp$ only, i.e. $\mathcal{T}(\varepsilon_\perp)$; and (iv) the condition of $\varepsilon_F \gg d_F$ for metallic bulk field emitter. 

For 2D materials, all of the above assumptions become invalid. Firstly, due to the lack of crystal periodicity along the $\perp$-direction, $\varepsilon_\perp$ does not follow a continuous dispersion relation. Instead, the cross-plane dynamics is described by discrete bound state energy levels as the electrons are confined by the surface potential \cite{vega} [see Fig. 1(c)]. In this case, the energy and wavevector partition rule in Eq. (\ref{part}) should be replaced by:
\begin{eqnarray}
\varepsilon\left(\mathbf{k}\right) &=& \varepsilon_\parallel \left(\mathbf{k}_\parallel\right) + \tilde{\varepsilon}_\perp (\tilde{\mathbf{k}}_\perp), \nonumber \\
\mathbf{k} &=& \mathbf{k}_\parallel + \tilde{\mathbf{k}}_\perp,
\end{eqnarray}
where $\tilde{\varepsilon}_\perp$ and $\tilde{\mathbf{k}}_\perp$ denotes the discrete bound state energy and the quantized $\perp$-directional wavevector, respectively. Secondly, the in-plane energy dispersion, $\varepsilon_\parallel$, is non-parabolic in $\mathbf{k}_\parallel$ for many 2D materials such as graphene and few-layer transition metal dichalcogenide (TMD). Thirdly, for the out-of-plane tunnelling of 2D electron gas into 3D bulk material, the inevitable presence of electron-electron and electron-impurity scatterings leads to a non-conserving lateral momentum (NCLM) during the tunneling process \cite{meshkov}. Such NCLM emission mechanism has been previously proposed as a route of enhancing thermionic cooling efficiency in semiconductor-based superlattice \cite{vashaee, vashaee2, dwyer, jeong}. 
Recently, NCLM has been experimentally observed in the cross-plane interfacial charge transport TMD-based van der Waals heterostructure \cite{zhu}.
Under the NCLM framework, the transmission probability is modified into a function of total energy, $\varepsilon$, instead of $\varepsilon_\perp$, i.e. $\mathcal{T}(\varepsilon_\perp) \to \mathcal{T}(\varepsilon)$ and its WKB form is written as \cite{meshkov}
\begin{equation}
\mathcal{T}(\varepsilon) = \lambda \exp \left[ -\frac{2}{\hbar} \int dz \sqrt{ 2m\left( V(z) - \varepsilon \right) } \right] ,
\end{equation}
where $\lambda$ is a parameter describing the strength of NCLM scattering processes, $V(z)$ is a $z$-dependent potential barrier and the emission occurs along the $z$-direction. 
The actual value of $\lambda$ varies depending on the quality of field emitter such as impurities and structural defects \cite{chen_NP}.
Furthermore, the substrates can also affect $\lambda$ via the dielectric screening effect \cite{ando, hwang, ponomarenko}.
Therefore, the magnitude of the SFE current in 2D material field emitter is expected to vary significantly in samples of different qualities and/or substrate configurations.
Using the experimental data of graphene-based Schottky contact \cite{sinha}, we estimate $\lambda = 10^{-4}$ for graphene electron emitter and this value is used throughout the following discussions (see Sec. V.A below for detailed discussion). 
Lastly, the assumption of $\varepsilon_F \gg d_F$ is no longer warranted for zero-gap 2D materials such as graphene in which the $\varepsilon_F$ can be electrostatically tuned from 0 to $0.9$ eV \cite{chen_FL}, thus allowing the condition of $\varepsilon_F \approx d_F$ or even $\varepsilon_F \ll d_F$.

The emission current density in 2D materials can be constructed by replacing the continuous $\varepsilon_\perp$-integral in Eq. (\ref{gen_J}) with a summation of all $i$ quantized bound states ($i = 1,2,3 \cdots$), denoted as $\varepsilon_{\perp, i}$, below the surface potential, i.e.
\begin{eqnarray}\label{C}
\mathcal{J}_{\text{2D}}^{\left( \text{C} \right)} &=& \frac{e}{L_\perp}\sum_i v_\perp(\varepsilon_{\perp,i}) \mathcal{T}(\varepsilon_{\perp,i}) \nonumber \\
&& \times \int_0^{\infty} D(\varepsilon_\parallel) d\varepsilon_\parallel f(\varepsilon_\parallel + \varepsilon_{\perp,i}),
\end{eqnarray}
\begin{eqnarray}\label{NC}
\mathcal{J}_{\text{2D}}^{\left( \text{NC} \right)} &=& \frac{e}{L_\perp}\sum_i v_\perp(\varepsilon_{\perp,i})  \nonumber \\
 &\times&  \int_0^{\infty} D(\varepsilon_\parallel) d\varepsilon_\parallel \mathcal{T}(\varepsilon_\parallel + \varepsilon_{\perp,i}) f(\varepsilon_\parallel + \varepsilon_{\perp,i}), \nonumber \\
\end{eqnarray}
where the superscript $(\text{C})$ and $(\text{NC})$ denotes CLM and NCLM models, respectively. 
The term $v_\perp(\tilde{\varepsilon}_{\perp, i}) = \sqrt{2\tilde{\varepsilon}_{\perp, i}/m}$ is the cross-plane velocity. 
In Eqs. (\ref{C}) and (\ref{NC}), the 2D material thickness, $L_\perp \approx 0.335$ nm for graphene, arises from the discrete summation $L_\perp^{-1} \sum_{k_\perp}$ [see Eq. (\ref{J})].
For single layer graphene, the energy difference between the Dirac point and the vacuum level is reported as $\Phi_{B0} \approx 4.5$ eV \cite{yu_wf, xu_wf, chris} (Note that $\Phi_B = \Phi_{B0} + \varepsilon_F$). 
We thus assume that the SFE process is chiefly contributed by only one bound state level lying at $4.5$ eV below the vacuum level. 
This bound state energy, denoted as $\varepsilon_{\perp, 1} \equiv \tilde{\varepsilon}_{\perp}$, is numerically solved from the time-independent Schr\"odinger's equation using a finite square well model proposed by Vega and de Abajo \cite{vega} [see Fig. 1(c)]. By matching the wavefunction maxima of the 2$p_z$ orbital of graphene and fitting the bound state energy to $\Phi_{B0} = 4.5$ below the vacuum level, the quantum well width, $L_0$, is numerically calculated to be 0.12 nm \cite{vega2} and the quantum well depth is $E_0 = 42.8$ eV, which yields the bound state energy of $\tilde{\varepsilon}_{\perp} = E_0 - \Phi_{B0} = 38.3$ eV [see Fig. 1(c) for energy band diagram]. We set this value of $\tilde{\varepsilon}_{\perp}$ as the zero-reference of total energy, i.e $ \tilde{\varepsilon}_{\perp} + \varepsilon_\parallel \to \varepsilon_\parallel$ thus allowing the emission current density to be simplified as
	\begin{subequations}\label{J_gen}
	\begin{equation}\label{J_gen_2D_C}
	\mathcal{J}_{\text{2D}}^{\left( \text{C} \right)} = \frac{ e v_\perp(\tilde{\varepsilon}_{\perp}) \mathcal{T}(\tilde{\varepsilon}_\perp) }{L_\perp}  \int_0^{\infty} D(\varepsilon_\parallel) d\varepsilon_\parallel  f(\varepsilon_\parallel),
	\end{equation}
	\begin{equation}\label{J_gen_2D_NC}
	\mathcal{J}_{\text{2D}}^{\left( \text{NC} \right)} = \frac{ ev_\perp(\tilde{\varepsilon}_{\perp}) }{L_\perp}  \int_0^{\infty} D(\varepsilon_\parallel) d\varepsilon_\parallel \mathcal{T}(\varepsilon_\parallel) f(\varepsilon_\parallel).
	\end{equation}
	\end{subequations}
A key difference between CLM and NCLM models is that for $\mathcal{J}_{\text{2D}}^{\left( \text{C} \right)} $, the $\mathcal{T}(\tilde{\varepsilon}_\perp)$ can be factored out from the $\varepsilon_\parallel$-integral since it is a function of solely $\tilde{\varepsilon}_\perp$ while for $\mathcal{J}_{\text{2D}}^{\left( \text{NC} \right)} $, the $\mathcal{T}(\varepsilon_\parallel)$ remains implicit to the $\varepsilon_\parallel$-integral [Fig. 1(d)].

\section{Theory of surface field emission from 2D materials}

\begin{figure}[b]
	\includegraphics[scale=0.45]{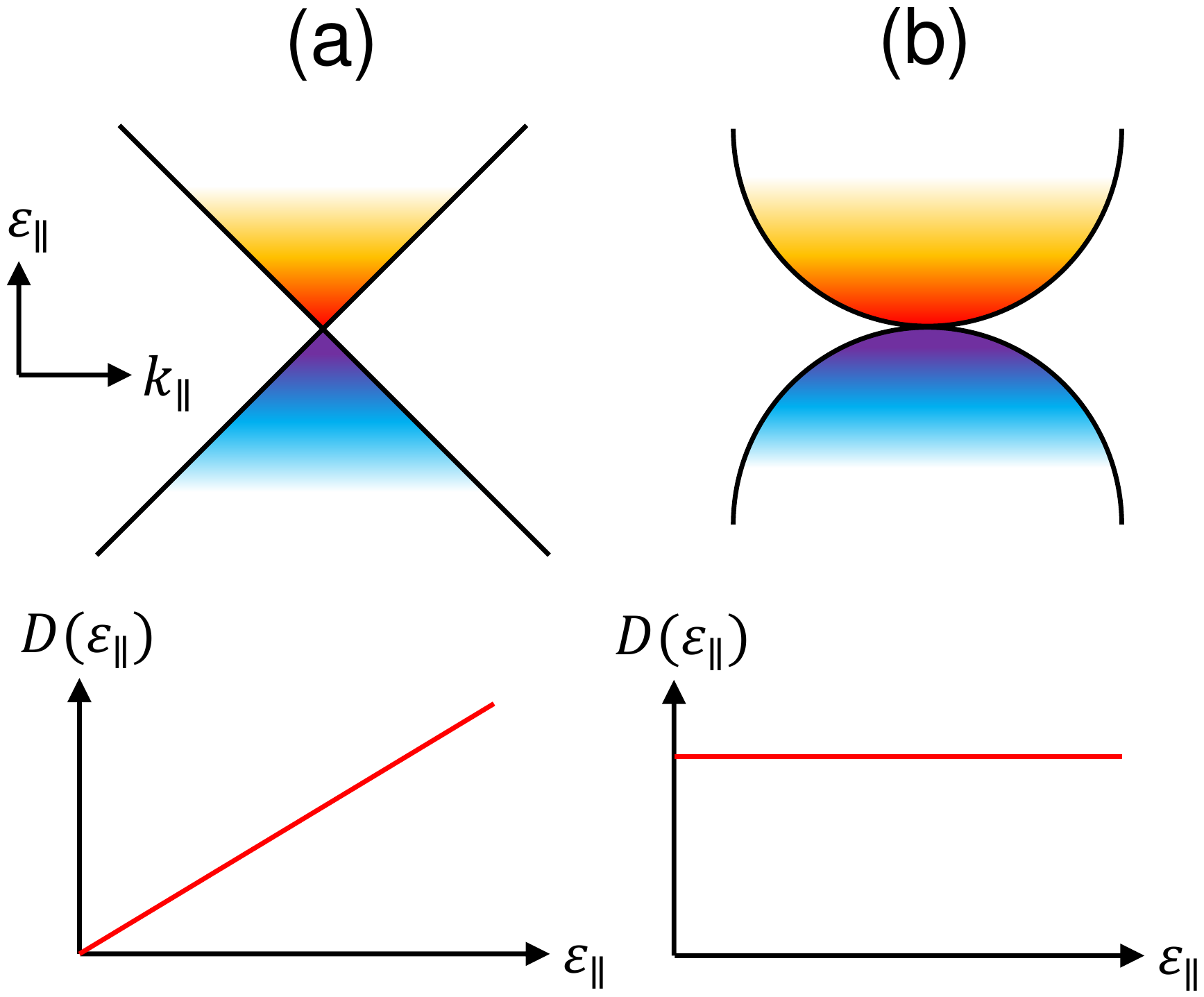}
	\caption{The energy dispersion and the lateral DOS of (a) $\varepsilon_\parallel \propto k_\parallel$ in monolayer graphene and (b) $\varepsilon_\parallel \propto k_\parallel^2$ in bilayer graphene.}
\end{figure}

\subsection{Lateral energy dispersion and density of states in 2D materials}

In the following, Eqs. (\ref{J_gen_2D_NC}) and (\ref{J_gen_2D_C}) are solved for two major classes of $\varepsilon_\parallel$, namely the non-relativistic parabolic energy dispersion of $\varepsilon_\parallel = \hbar^2 k_\parallel^2 /2m^*$ and the linear gapless Dirac energy dispersion of $\varepsilon_\parallel = \hbar v_F k_\parallel$. 
The $k_\parallel$-linear energy dispersion describes the low energy electrons residing at the $K$ and $K'$ Dirac cone of graphene monolayer \cite{neto}. 
For Bernal-stacked bilayer graphene, the effective low energy dispersion around the $K$ and $K'$ points follows the $k_\parallel$-parabolic energy dispersion \cite{mccann}. The lateral DOS can be obtained as
\begin{subequations}\label{DOS}
	\begin{equation}
	D_{\text{2DEG}}(\varepsilon_\parallel) = \frac{g_{s,v} m^*}{2\pi \hbar^2},
	\end{equation}
	\begin{equation}
	D_{\text{Dirac}}(\varepsilon_\parallel) = \frac{g_{s,v} \varepsilon_\parallel}{2\pi \hbar^2 v_F^2},
	\end{equation}
\end{subequations}
respectively, for parabolic and linear dispersion of $\varepsilon_\parallel$. 
Here, $v_F \approx 10^6$ m/s is the Fermi velocity in monolayer graphene \cite{neto} and $m^* \approx 0.03m$ \cite{li} is the electron effective mass in Bernal-stacked bilayer graphene \cite{li}.
The linear and parabolic form of $\varepsilon_\parallel$ and $D(\varepsilon_\parallel)$ are shown in Fig. 2. 

For few-layer black phosphorus field emitter \cite{suryawanshi}, the lateral energy dispersion near the conduction band edge can be approximated by an \emph{anisotropic-parabolic} spectrum \cite{jiang, han}, i.e. $\varepsilon_\parallel(k_x, k_y) \approx \hbar^2 k_x^2/2m_{cx}^* + \hbar^2 k_y^2/2m_{cy}^*$, where $\mathbf{k}_\parallel = (k_x, k_y)$, and $m_{cx,cy}^*$ is the conduction band electron effective mass in $x$- and $y$-directions, respectively. The lateral DOS follows the same form as the isotropic parabolic lateral DOS in Eq. (\ref{DOS}), i.e. $D_{\text{BP}}(\varepsilon_\parallel) = g_{s}m_d/2\pi\hbar^2$, where $m_d \equiv \sqrt{m_{cx}^*m_{cy}^*}$ is the DOS electron effective mass and $g_s$ is the spin degeneracy. Thus, the FE models developed below can be readily generalized for few-layer black phosphorus field emitter by substituting $m^* \to m_d$ and $g_{s,v} \to g_{s}$.

\subsection{2D material field emission with conserving lateral momentum (CLM)}

For SFE from 2D material with CLM, the WKB transmission probability is given as $\mathcal{T}(\tilde{\varepsilon}_\perp) = D_F \exp\left[ \left(\tilde{\varepsilon}_\perp - \varepsilon_F\right)/d_F \right]$, where $D_F$ is defined in Eq. (\ref{FE_cond}). By setting $\tilde{\varepsilon}_\perp$ as the zero-reference of $\varepsilon_F$, i.e. $ (\varepsilon_F+\tilde{\varepsilon}_\perp) \to \varepsilon_F$ and employing the $T=0$ K condition, Eq. (\ref{J_gen_2D_C}) can be solved as
\begin{subequations}
\begin{equation}\label{C_FE}
\mathcal{J}_{  \text{2DEG} }^{(\text{C})} 
= \frac{v_\perp(\tilde{\varepsilon}_\perp)}{L_\perp} \frac{g_{s,v} e m^* }{2\pi \hbar^2}\varepsilon_F
D_F \exp \left( -\frac{\varepsilon_F}{d_F} \right) ,
\end{equation}
\begin{equation}\label{C_FE}
\mathcal{J}_{ \text{Dirac} }^{(\text{C})}
= \frac{v_\perp(\tilde{\varepsilon}_\perp)}{L_\perp} \frac{g_{s,v} e}{2\pi \hbar^2v_F^2 }
\frac{\varepsilon_F^2}{2}
D_F \exp \left( -\frac{\varepsilon_F}{d_F} \right) ,
\end{equation}
\end{subequations}
respectively, for 2DEG and Dirac energy dispersions.
Although both traditional FN law for bulk material and the modified FE model for 2D material in Eq. (\ref{C_FE}) belong to the same CLM models, their pre-factor exhibits contrasting $F$-dependence -- the pre-factor in Eq. (\ref{C_FE}) is $F$-independent rather than the $F^2$-dependence as in the case of traditional FN law for 3D bulk material. 
Accordingly, the FN-plot is modified to 
\begin{subequations}\label{CLM_FN_plot}
\begin{equation}
\ln\left(\mathcal{J}_{  \text{2DEG} }^{(\text{C})} \right) \propto - \frac{1}{F}
\end{equation}
\begin{equation}
\ln\left(\mathcal{J}_{  \text{Dirac} }^{(\text{C})} \right) \propto - \frac{1}{F}
\end{equation}
\end{subequations}
This is in stark contrast to the traditional FN-plot of $\ln(\mathcal{J}/F^2) \propto -1/F$.
Furthermore, the FE current density exhibits a Fermi level scaling of $\mathcal{J}_{\text{2DEG}}^{(\text{C})} \propto \varepsilon_F$ and $\mathcal{J}_{\text{Dirac}}^{(\text{C})} \propto \varepsilon_F^2$. 
The FE current density vanishes as $\varepsilon_F \to 0$ due to the depletion of electrons.

It is important to point out the following: due to the absence of $F$-dependence in the pre-factor, the surface field emission current density from 2D material will saturate at high-field limit of $d_F \gg \varepsilon_F$ when both $D_F$ and $\exp(-\varepsilon_F/d_F)$ approaches unity.
In this case, the \emph{saturated field emission current density} is
\begin{subequations}\label{C_small}
\begin{equation}
\mathcal{J}_{\text{2DEG} \star }^{(\text{C})} = \frac{v_\perp(\tilde{\varepsilon}_\perp)}{L_\perp} \frac{g_{s,v} e m^*}{2\pi \hbar^2}\varepsilon_F
\end{equation}
\begin{equation}
\mathcal{J}_{ \text{Dirac} \star }^{(\text{C})} = \frac{v_\perp(\tilde{\varepsilon}_\perp)}{L_\perp} \frac{g_{s,v} e}{2\pi \hbar^2} \frac{\varepsilon_F^2}{2v_F^2}
\end{equation}
\end{subequations}
where the subscript `$\star$' denotes saturated emission. The $F$-independent pre-factor in Eq. (\ref{C_FE}) arises from the absence of $\varepsilon_\perp$-integral in the formulation of 2D material current density [see Eq. (\ref{J_gen})].
Thus, this \emph{saturated surface field emission} represents a direct consequence of the reduced dimensionality which can be used to probe the 2D nature of 2D-material-based field emitter.

\begin{figure*}[t]
	\includegraphics[scale=1]{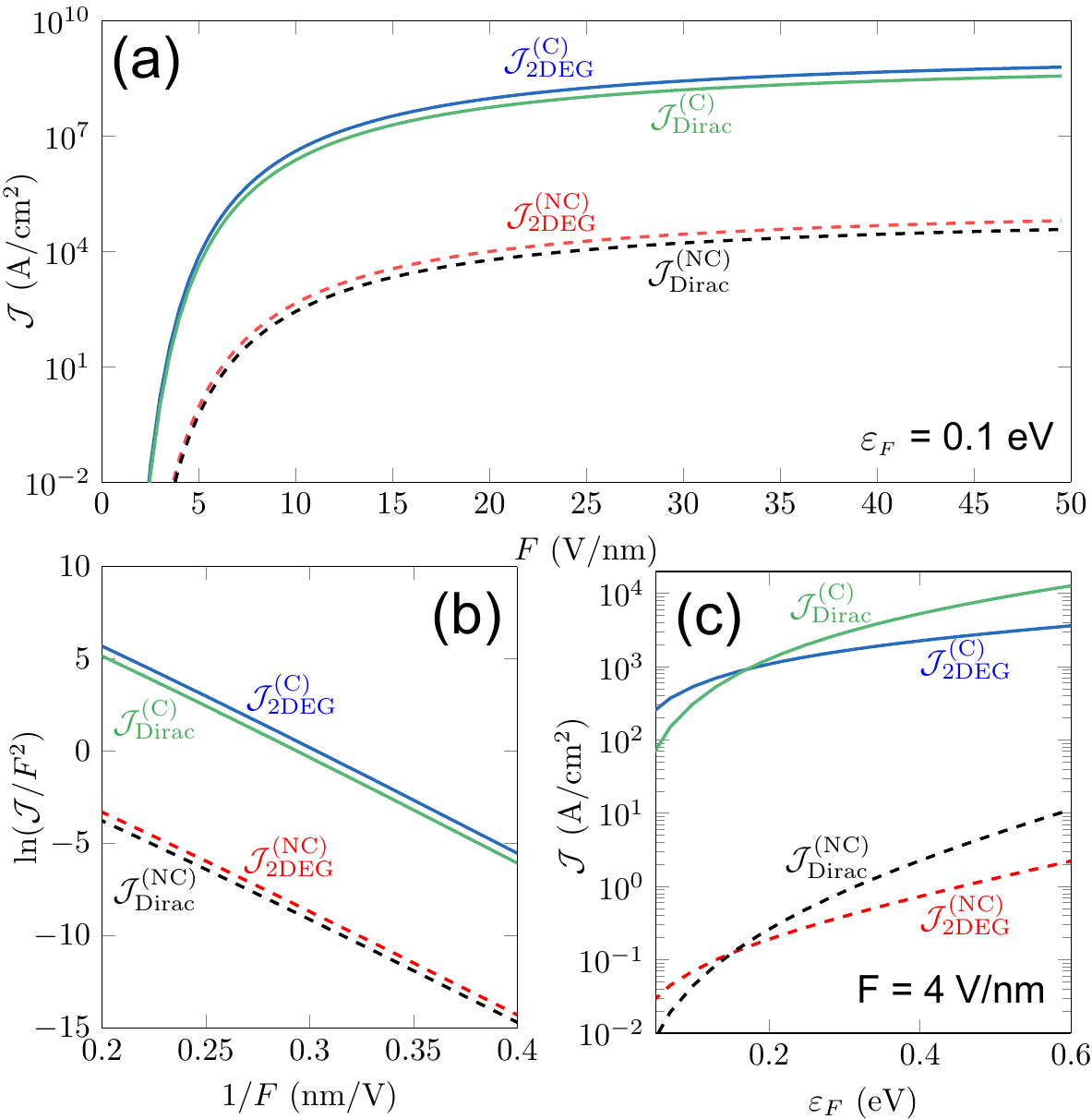}
	\caption{Field-induced surface electron emission current density characteristics for 2D material with parabolic and linear lateral energy dispersion. (a) $\mathcal{J}$-$F$ characteristics. (b) Traditional FN plots, $\ln(\mathcal{J}/F^2)$ versus $1/F$; (c) $\varepsilon_F$-dependence of $\mathcal{J}$. In (a) and (b), the Fermi level is $\varepsilon_F = 0.1$ eV. Note that the following material-dependent parameters are used throughout the manuscript: $v_F = 10^6$ m/s, $m^* = 0.03 m$, $L_\perp = (0.335, 0.670)$ nm respectively for linear (single layer) and parabolic (bilayer graphene) lateral energy dispersion, $\tilde{\varepsilon}_\perp = 38.3$ eV, $g_{s,v} = 4$, $\lambda = 10^{-4}$ and $\Phi_B \equiv \Phi_{B0} + \varepsilon_F$ where $\Phi_{B0} = 4.5$ eV.}
\end{figure*}

\subsection{2D material field emission with non-conserving lateral momentum (NCLM)}

In the case of NCLM, the WKB transmission probability across a triangular surface can be written as a function of \emph{total energy} \cite{meshkov}, i.e. $\mathcal{T}(\varepsilon) = \lambda D_F\exp \left( \frac{\varepsilon - \varepsilon_F}{d_F} \right)$. 
At $T=0$ K, Eq.(\ref{J_gen_2D_NC}) can be analytically solved as
\begin{subequations}\label{NC_FE}
	\begin{eqnarray}
	\mathcal{J}_{\text{2DEG}}^{(\text{NC})} &=& \lambda\frac{v_\perp(\tilde{\varepsilon}_\perp)}{L_\perp} \frac{g_{s,v} m^* e}{2\pi \hbar^2}  \nonumber \\
	&& \times D_F \left[ 1 - \exp\left(-\frac{\varepsilon_F}{d_F}\right) \right] d_F ,
	\end{eqnarray}
	\begin{eqnarray}
	\mathcal{J}_{\text{Dirac}}^{(\text{NC})} &=& \lambda\frac{v_\perp(\tilde{\varepsilon}_\perp)}{L_\perp} \frac{g_{s,v} e}{2\pi \hbar^2 v_F^2}  \nonumber \\ 
	&& \times D_F \left[ \frac{\varepsilon_F}{d_F} + \exp\left(-\frac{\varepsilon_F}{d_F}\right)- 1 \right] d_F^2 .
	\end{eqnarray}
\end{subequations}
In the low-field regime of $d_F \ll \varepsilon_F$ ($d_F \propto F$), Eq. (\ref{NC_FE}) reduces to
\begin{subequations}\label{J_ef_gg_df}
\begin{eqnarray}
\mathcal{J}_{\text{2DEG}}^{(\text{NC})} (d_F \ll \varepsilon_F) &=&
\lambda\frac{v_\perp(\tilde{\varepsilon}_\perp)}{L_\perp} \frac{g_{s,v}em^* }{2\pi\hbar^2} \nonumber \\
&& \times D_F d_F \exp \left( -\frac{\varepsilon_F}{d_F} \right),
\end{eqnarray}
\begin{eqnarray}
\mathcal{J}_{\text{Dirac}}^{(\text{NC})} (d_F \ll \varepsilon_F) &=& \lambda\frac{v_\perp(\tilde{\varepsilon}_\perp)}{L_\perp} \frac{g_{s,v}em^* }{2\pi\hbar^2} \nonumber \\
&& \times \frac{ \varepsilon_F}{v_F^2} D_F d_F \exp \left( -\frac{\varepsilon_F}{d_F} \right),
\end{eqnarray}
\end{subequations}
where the pre-factor is linear in $F$ for both 2DEG and Dirac dispersions. Correspondingly, the FN-plot for NCLM model at low-field regime is modified as
\begin{subequations}
\begin{equation}
\ln\left[\frac{\mathcal{J}_{  \text{2DEG} }^{(\text{NC})}(d_F \ll \varepsilon_F)}{F} \right] \propto - \frac{1}{F},
\end{equation}
\begin{equation}
\ln\left[\frac{\mathcal{J}_{  \text{Dirac} }^{(\text{NC})}(d_F \ll \varepsilon_F)}{F} \right] \propto - \frac{1}{F}.
\end{equation}
\end{subequations}
In the high-field regime of $d_F \gg \varepsilon_F$, the FE current density becomes equivalent to Eq. (\ref{C_FE}) apart from an additional factor of $\lambda$, 
\begin{subequations}\label{NC_small}
\begin{equation}
\mathcal{J}_{\text{2DEG}}^{(\text{NC})} (d_F \gg \varepsilon_F) = \lambda \mathcal{J}_{\text{2DEG}}^{(\text{C})}(d_F \gg \varepsilon_F),
\end{equation}
\begin{equation}
\mathcal{J}_{\text{Dirac}}^{(\text{NC})} (d_F \gg \varepsilon_F) = \lambda \mathcal{J}_{\text{Dirac}}^{(\text{C}) } (d_F \gg \varepsilon_F),
\end{equation}
\end{subequations}	
The modified FN-plot for NCLM model thus follow the same relation as Eq. (\ref{CLM_FN_plot}) at high-field regime. At the limit of very high applied electric field strength, the FE current densities for CLM and NCLM converge to the same saturated value given by
\begin{subequations}\label{NC_small}
\begin{equation}
\mathcal{J}_{\text{2DEG} \star}^{(\text{NC})} = \lambda \mathcal{J}_{\text{2DEG}  \star}^{(\text{C})} 
\end{equation}
\begin{equation}
\mathcal{J}_{\text{Dirac} \star}^{(\text{NC})} = \lambda \mathcal{J}_{\text{Dirac}  \star}^{(\text{C})}
\end{equation}
\end{subequations}	
The modified FN-plot in Eq. (\ref{CLM_FN_plot}) and the saturated emission current densities in Eqs. (\ref{C_small}) and (\ref{NC_small}) are universal high-field features for both CLM and NCLM models.
The convergence of NCLM into CLM models at $d_F \gg \varepsilon_F$ can be explained as follow. When $d_F$ is sufficiently large, the electrons contributing to FE has a relatively small lateral energy component, $\varepsilon_\parallel$, in comparison with $d_F$. In this case, the NCLM transmission probability becomes nearly $\varepsilon_\parallel$-independent, i.e. $\mathcal{T}(\tilde{\varepsilon}_\perp + \varepsilon_\parallel) \approx \mathcal{T}(\tilde{\varepsilon}_\perp)$. 
As a result, the NCLM model [Eq. (\ref{NC_FE})] converges to the CLM model [Eq. (\ref{C_FE})].

\subsection{Field emission characteristics of CLM and NCLM models and comparison with experiment}

\begin{figure}
	\includegraphics[scale=.65]{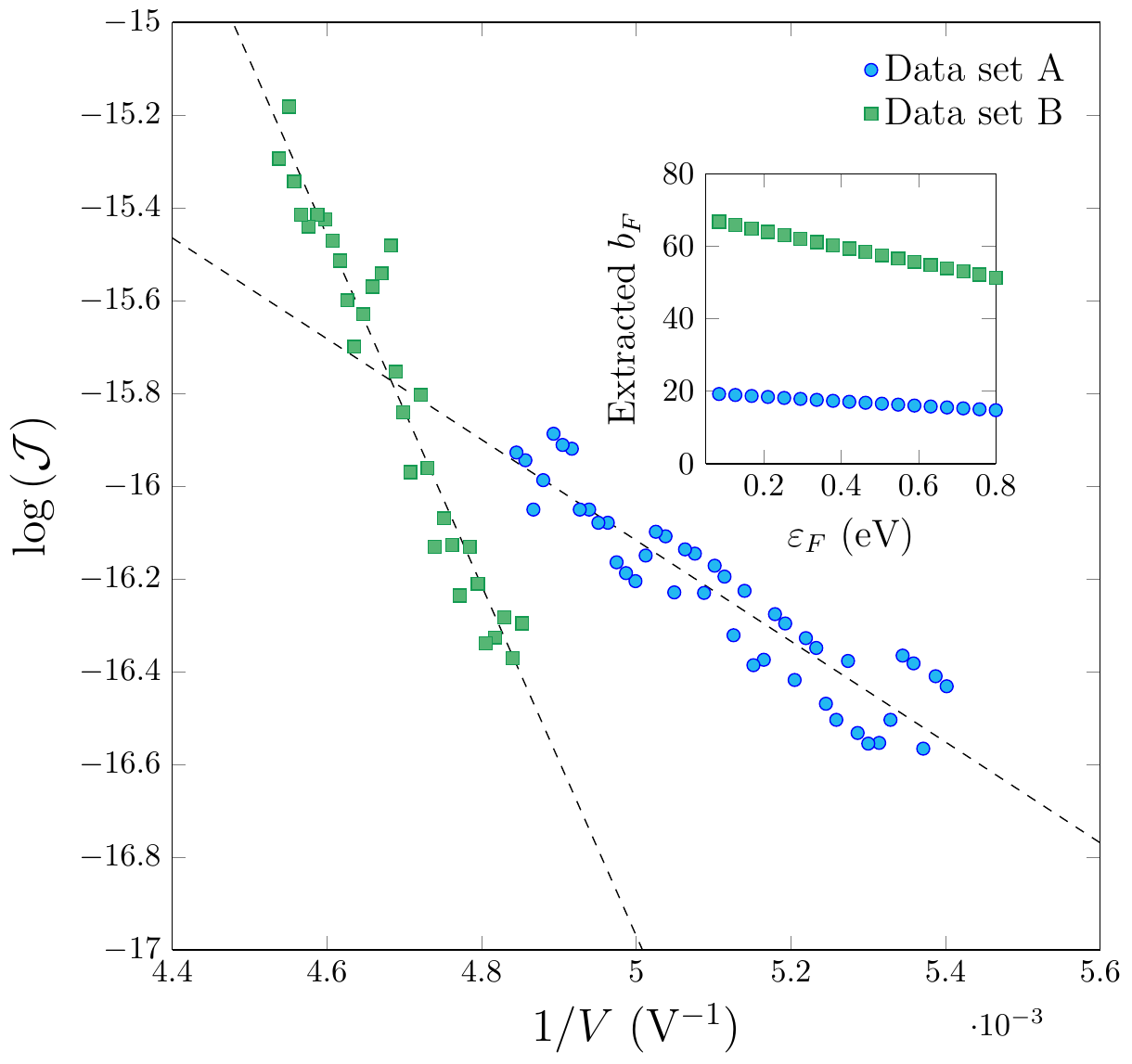}
	\caption{Modified FN-plot, $\ln(\mathcal{J})$ vs. $1/V$, of surface field emission from suspended graphene (experimental data from Ref.\cite{xu}. Inset shows the field enhancement factor, $b_F$, extracted using Eq. (\ref{field_enhancement}) with $\Phi_{B0} = 4.5$ eV. }
\end{figure}

In Figs. 3(a) and (b), the FE characteristics of CLM and NCLM models are shown with $\varepsilon_F = 0.1$ eV. In Fig. 3(a), the required electric field strength to achieve $0.01$ A/cm$^2$ of emission current density is typically $F \approx 2.5$ V/nm and $F \approx 4$ V/nm, respectively, for CLM and NCLM models. 
At high-field regime, the emission current densities gradually saturate. For graphene, the saturated emission current density is in the order of $\sim 10^{4}$ A/cm$^2$ and $\sim 10^{8}$ A/cm$^2$, respectively, for NCLM and CLM models, which is well below the breakdown current density experimentally determined in suspended graphene, i.e. $ \mathcal{J}_{\text{break} } > 10^{12}$ A/cm$^2$ \cite{gruber}.
Hence, single layer graphene may serve as a feasible platform to experimentally probe the saturated field emission effect. 
The traditional FN plot is shown in Fig. 3(b). 
Due to the strong dominance of the exponential term, $D_F$, the FN plots can still be fitted by straight lines although the pre-factor of the revised FE model has a completely different $F$-dependence in comparison with the traditional FN model.  
In Fig. 3(c), the $\varepsilon_F$-dependence of the emission current density is plotted with $F=4$ V/nm. 
The FE current densities increase monotonously as $\varepsilon_F$ increases. At low $\varepsilon_F$, we have $\mathcal{J}_{\text{2DEG}} > \mathcal{J}_{\text{Dirac}}$. The opposite case of $\mathcal{J}_{\text{Dirac}} > \mathcal{J}_{\text{2DEG}}$ is obtained at $\varepsilon_F \gtrsim 0.1$ eV. This cross-over occurs since $\mathcal{J}_{\text{Dirac}}$ is approximately one order of $\varepsilon_F$ higher than $\mathcal{J}_{\text{2DEG}}$ for both CLM and NCLM models [see Eqs. (\ref{C_FE}) and (\ref{NC_FE})].

In Fig. 4, the high-field segments of the graphene SFE experimental data reported in Ref. \cite{xu} is analysed using the proposed \emph{modified FN-plot} in Eq. (\ref{CLM_FN_plot}). Both modified FN-plot and the traditional FN plot produce reasonably good linear fit. Nonetheless, the \emph{coefficient of determination} (COD), which quantifies the goodness of a linear fit (COD $=1$ represents perfect straight line), is consistently improved for both sets of experimental data when fitted using the modified FN plot -- the CODs are improved from 0.813 to 0.866 (Data set A in Fig. 4) and from 0.927 to 0.934 (Data set B in Fig. 4).

The field enhancement factor, $b_F$, can be extracted from the modified FN-plot via
\begin{equation}\label{field_enhancement}
b_F = D\frac{B_{FN}\left(\Phi_{B0}-\varepsilon_F\right)^{3/2}\left(1 + \frac{3}{2}\frac{\varepsilon_F}{\Phi_{B0} - \varepsilon_F}\right)}{\mathcal{M}},
\end{equation}
where $\Phi_{B0}$ is the energy difference between Dirac point (conduction band bottom) and vacuum level for linear (parabolic) lateral energy dispersion. $\mathcal{M}$ is the slope extracted from the modified FN-plot. Here, we have assumed a simple field enhancement of $F = b_FV/D$ where $D$ is the anode-cathode separation and $V$ is the applied voltage. Since the value of $\varepsilon_F$ is not known, the extracted $b_F$ is plotted in the inset of Fig. 4 for a range of $\Phi_{B0} - \varepsilon_F$. For both sets of FE experimental data, $b_F = 20 \sim 60$ is extracted. This low magnitude of field enhancement factor is consistent with the fact that sharp tips and edges are absent in graphene SFE configuration.

\subsection{Finite-temperature surface field emission from 2D materials}

We now generalized the SFE models for $T > 0$ K.
Because $\Phi_B \gg k_BT$ at $T = 300$ K, we can safely neglect the TE effect.  
The finite-temperature CLM model of field emission from 2D materials can be straightforwardly solved as,
\begin{widetext}
\begin{subequations}
\begin{equation}
\mathcal{J}_{\text{2DEG}}^{(\text{C})} (T) =
\frac{ev_\perp(\tilde{\varepsilon}_\perp) }{L_\perp} \frac{g_{s,v}em^*}{2\pi\hbar^2} k_BT \ln\left[\exp\left( -\frac{\varepsilon_F}{k_BT} \right)+ 1\right] D_F \exp \left( - \frac{\varepsilon_F}{d_F} \right) ,
\end{equation}
\begin{equation}
\mathcal{J}_{\text{Dirac}}^{(\text{C})} (T) =
\frac{ev_\perp(\tilde{\varepsilon}_\perp) }{L_\perp} \frac{g_{s,v}e}{2\pi\hbar^2v_F^2}
\left(k_BT\right)^2 \mathcal{F}_1\left(\frac{\varepsilon_F}{k_BT}\right)  D_F \exp \left( - \frac{\varepsilon_F}{d_F} \right) ,
\end{equation}
\end{subequations}
where $\mathcal{F}_1(x) = \int_0^\infty dt \left[\exp( t-x ) + 1 \right]^{-1}$ is the first-order complete Fermi-Dirac integral \cite{abramowitz}.
For NCLM model, the finite-temperature model can be semi-analytically expressed in terms of generalized hypergeometric function \cite{abramowitz}, $_p\mathscr{F}_q(z)$ where $p$ and $q$ are positive integers. We obtain,
\begin{subequations}
\begin{eqnarray}\label{FE_NC_T_2DEG}
	\mathcal{J}_{\text{2DEG}}^{(\text{NC})}(T) 
	&=& \frac{ev_\perp(\tilde{\varepsilon}_\perp) }{L_\perp} \frac{g_{s,v}em^*}{2\pi\hbar^2} D_F d_F \sum_{i=1}^{2} f_i(\gamma, \varepsilon_f, d_F),
\end{eqnarray}
\begin{eqnarray}\label{FE_NC_T_dirac}
	\mathcal{J}_{\text{Dirac}}^{(\text{NC})}(T)
	&=& \frac{ev_\perp(\tilde{\varepsilon}_\perp) }{L_\perp} \frac{g_{s,v}e}{2\pi\hbar^2v_F^2} D_F d_F^2 \sum_{i=1}^{3} g_i(\gamma, \varepsilon_f, d_F),
\end{eqnarray}
\end{subequations}
where $\gamma \equiv k_BT/d_F$. Note that in deriving Eq. (\ref{FE_NC_T_2DEG}), we have set the upper limit of $\varepsilon_\parallel$-integral to $2\Phi_B/3$ to exclude the over-barrier thermionic emission. The terms $f_i$ and $g_i$ are given as functions of $_p\mathscr{F}_q(z)$:
	\begin{equation*}
	f_1(\beta, \varepsilon_f, d_F) = \exp \left( \frac{2\Phi_B/3 - \varepsilon_F}{d_F} \right) {_2\mathscr{F}_1} \left(1, \gamma, 1 + \gamma, -e^{\frac{2\Phi_B/3 - \varepsilon_F }{k_BT} } \right),
	\end{equation*}
	\begin{equation*}
	f_2(\beta, \varepsilon_f, d_F) = - \exp \left( - \frac{\varepsilon_F}{d_F} \right) {_2\mathscr{F}_1} \left(1, \gamma, 1 + \gamma, -e^{\frac{\varepsilon_F }{k_BT} } \right) ,
	\end{equation*}
	\begin{equation*}
	g_1(\gamma, \varepsilon_f, d_F) = \frac{2\Phi_B}{3d_F} \exp \left( \frac{2\Phi_B/3 - \varepsilon_F}{d_F} \right)  {_2\mathscr{F}_1} \left(1, \gamma, 1 + \gamma, -e^{\frac{2\Phi_B/3 - \varepsilon_F }{k_BT} } \right) ,
	\end{equation*}
	\begin{equation*}
	g_2(\gamma, \varepsilon_f, d_F) = - \exp \left( \frac{2\Phi_B/3 - \varepsilon_F}{d_F} \right) {_3\mathscr{F}_2 }\left( \{1, \gamma, \gamma\},\{1+\gamma, 1+\gamma\}, -e^{\frac{2\Phi_B/3 - \varepsilon_F }{k_BT} } \right),
	\end{equation*}
	\begin{equation}
	g_3(\gamma, \varepsilon_f, d_F) =  \exp \left( - \frac{\varepsilon_F}{d_F} \right)  {_3\mathscr{F}_2 }\left( \{1, \gamma, \gamma\},\{1+\gamma, 1+\gamma\}, -e^{- \frac{ \varepsilon_F }{k_BT} } \right).
	\end{equation}
\end{widetext}

\begin{figure}
	\includegraphics[scale=0.725]{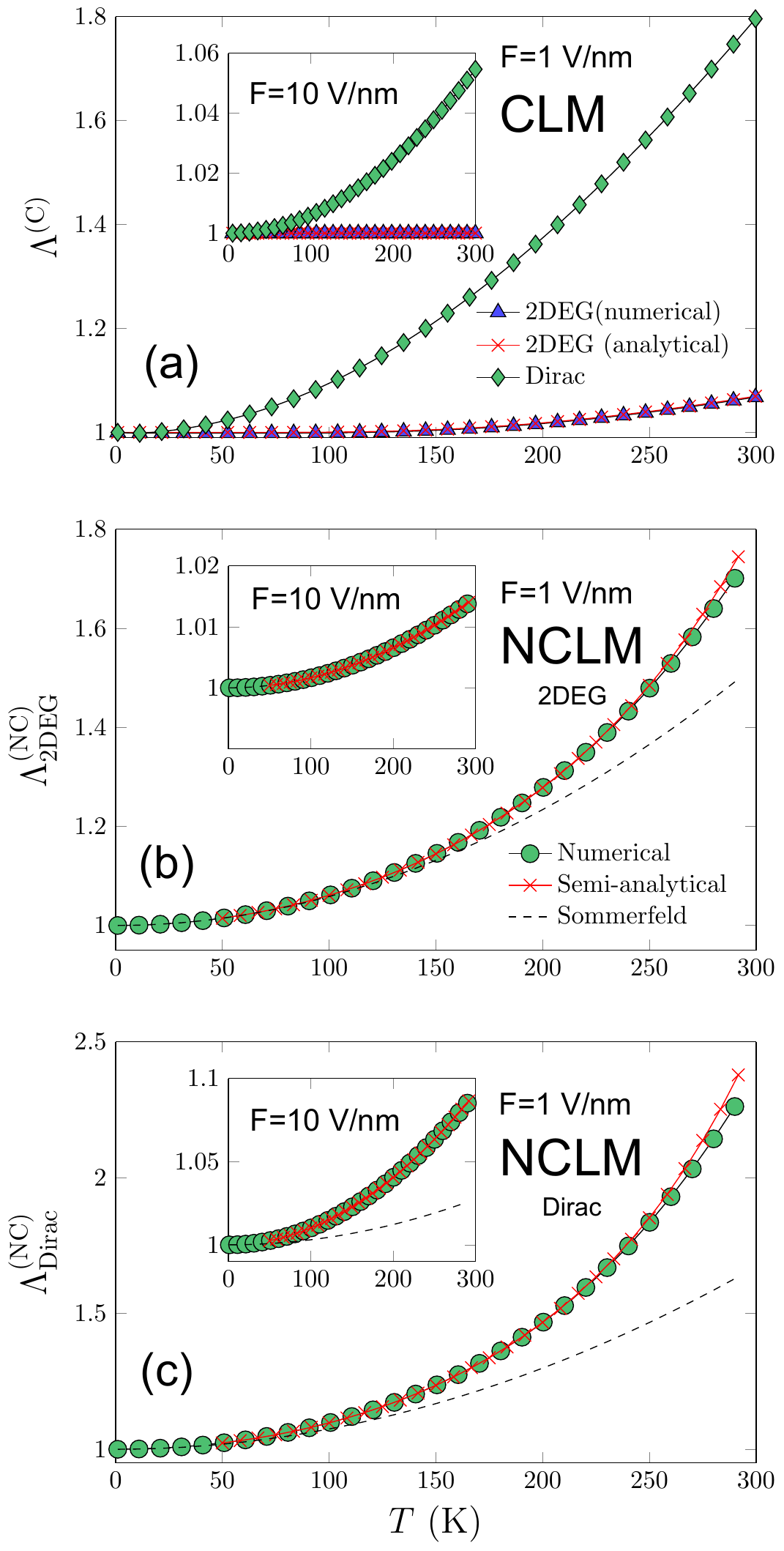}
	\caption{Thermal-enhancement of field emission current density at $T > 0$ K. (a) CLM model; (b) NCLM model with parabolic energy dispersion; and (c) NCLM model with linear energy dispersion. The main plots are calculated with $F = 1 V/nm$. The insets show the $\Lambda$'s with $F = 10$ V/nm. (Note that the semi-analytical results is evaluated for $T>50$ K due to the much reduced accuracy of the numerically-generated hypergeometric function at small $T$.) }
\end{figure}

The temperature dependence of the NCLM model is not immediately obvious from the semi-analytical forms in Eqs. (\ref{FE_NC_T_2DEG}) and (\ref{FE_NC_T_dirac}). To obtain simple analytical approximation of the $T>0$ NCLM model, we employ the Sommerfeld expansion \cite{foot_1} which is well-suited for the case of $k_BT \ll \varepsilon_F$ and $T\neq 0$. The finite temperature Sommerfeld correction to $\mathcal{J}^{(\text{NC})}$, i.e. $\Delta \mathcal{J}^{(\text{NC})}(T) \equiv \mathcal{J}^{(\text{NC})}(T) - \mathcal{J}^{(\text{NC})}(T=0) $, are
\begin{subequations}\label{SE}
\begin{equation}
\Delta {J}_{ \text{2DEG}}^{(\text{NC})} (T)=
\lambda \frac{g_{s,v} m^*e}{2\pi \hbar^2}\frac{\left(\pi k_BT\right)^2}{6} \frac{1}{d_F},
\end{equation}
\begin{equation}
\Delta \mathcal{J}_{ \text{Dirac}}^{(\text{NC})}(T)
=
\lambda \frac{g_{s,v}e}{2\pi \hbar^2 v_F^2}\frac{\left(\pi k_BT\right)^2}{6} 
 \frac{\varepsilon_F}{d_F}.
\end{equation}
\end{subequations}
In Fig. 5, we plot the thermal-enhancement ratio, defined as
\begin{equation}
\Lambda(T) \equiv \frac{\mathcal{J}(T)}{\mathcal{J}(T =0)},
\end{equation}
with $F = 1$ V/nm (main plots) and with $F = 10$ V/nm (insets). For both CLM and NCLM models, the thermal-enhancement of emission current is only profound at low-field regime. In Fig. 5(a), the numerical and semi-analytical results of $\Lambda_{\text{2DEG}}^{(\text{C})}$ and $\Lambda_{\text{Dirac}}^{(\text{C})}$ of CLM models are plotted. At $F=1$ V/nm, the room-temperature $\mathcal{J}^{(\text{C})}_{\text{Dirac}}(T)$ is thermally-enhanced by $\sim80$\% with respect to $\mathcal{J}^{(\text{C})}_{\text{Dirac}}(T=0)$ [Fig. 5(a)]. This is in stark contrast to the modest thermal-enhancement of $\sim6$\% at $F=10$ V/nm [inset of Fig 5(a)]. For parabolic energy dispersion, the thermal-enhancement of $\mathcal{J}^{(\text{C})}_{\text{2DEG}}$ is less dramatic, i.e. $<10$\% enhancement at $F=1$ V/nm and a nearly absence of thermal-enhancement at $F = 10$ V/nm.
The $\Lambda_{\text{2DEG}}^{(\text{NC})}$ and $\Lambda_{\text{Dirac}}^{(\text{NC})}$ of NCLM models are shown, respectively, in Figs. 5(b) and (c). 
The $\mathcal{J}_{\text{2DEG}}^{(\text{NC})}$ exhibit a similar thermal-enhancement of $\sim$80\% at $F = 1$ V/nm [Fig. 5(b)], but is suppressed to only $\sim 1.5$\% at $F = 10$ V/nm [inset Fig. 5(b)].
For $\mathcal{J}_{\text{Dirac}}^{(\text{NC})}$, a dramatic thermal-enhancement of $\sim 250$\% at room temperature is obtained with $F = 1$ V/nm [Fig. 5(c)] and is reduced to $<10$\% at $F=10$ V/nm [inset of Fig. 5(c)].

The room-temperature ($T = 300$ K) FE current densities can be ranked, respectively for $F = 1$ V/nm and for $F = 10$ V/nm, as
\begin{subequations}
	\begin{equation}
	\Lambda_{\text{Dirac}}^{(\text{NC})} > \Lambda_{\text{2DEG}}^{(\text{C})}  \approx \Lambda_{\text{2DEG}}^{(\text{NC})} > \Lambda_{\text{Dirac}}^{(\text{C})},
	\end{equation}
	\begin{equation}
	\Lambda_{\text{Dirac}}^{(\text{NC})} > \Lambda_{\text{2DEG}}^{(\text{C})}  >  \Lambda_{\text{2DEG}}^{(\text{NC})} > \Lambda_{\text{Dirac}}^{(\text{C})}.
	\end{equation}
\end{subequations}
Note that the semi-analytical solution based on hypergeometric functions is in good agreement with the full numerical result. Also, the $T^2$ thermal-enhancement predicted by the Sommerfeld expansion in Eq. (\ref{SE}) agrees well with the semi-analytical model up to $\sim 100$ K. 
The contrasting thermal-enhancement of FE current at low- and at high-field regimes can be qualitatively explained by the Sommerfeld expansion in Eq. (\ref{SE}).
As both $\Delta \mathcal{J}_{\text{2DEG}}^{(\text{NC})}$ and $\Delta \mathcal{J}_{\text{Dirac}}^{(\text{NC})}$ are proportional to $1 / d_F$, the thermal-enhancement becomes severely quenched at elevated electric field strength.

\section{Absence of space-charge-limited emission in 2D material field emitter}

\begin{figure}[t]
	\includegraphics[scale=0.25]{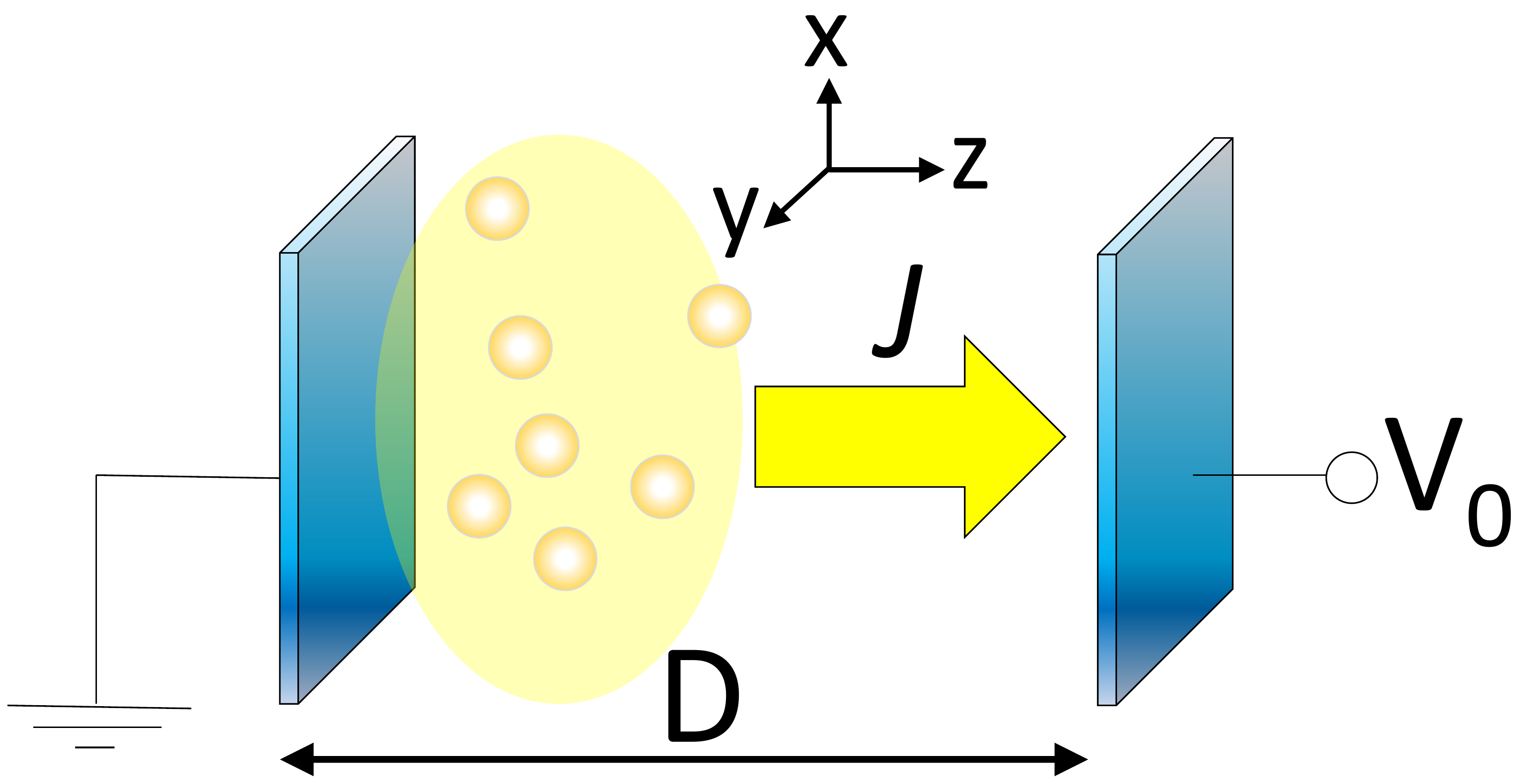}
	\caption{Schematic drawing of the space-charge-limited emission from 2D material field emitter.}
\end{figure}

We now construct a general emission model -- covering both FE and SCL regims -- through the combination of FE models developed above and Poisson equation \cite{lau}.
We consider a planar vacuum diode composed of two large area 2D-material-based electrodes as shown in Fig. 6.
We neglect the quantum SCL effect \cite{lau_QSCLC, ang_QSCLC, koh} which is significant in nanoscale vacuum gap when the electrode-separation is comparable to the de Broglie wavelength of the emitted electrons. 
We further neglect finite-temperature effect since it produces only a small correction to the emission current densities at high electric field regime as discussed above.
The electrostatic field distribution within the vacuum gap can be described by a 1D Poisson equation, $d^2V(z)/dz^2 = en(z,t)/\epsilon_0 = e\mathcal{J}/m\epsilon_0$, where the emission current density is $\mathcal{J} = n(z,t)ev_\perp(z,t)$. The electron is emitted along the $z$-direction. Here, $\epsilon_0$ is the permittivity of free space and $n(z,t)$ is the space charge density in the vacuum gap. By a change of variable, $z\to t$, and the definition of $v_\perp = dz/dt$, the Poisson equation can be transformed into the Llewellyn form \cite{lau},
\begin{equation}\label{PE}
\frac{d^2v_\perp(t)}{dt^2} = \frac{e\mathcal{J}}{m\epsilon_0},
\end{equation}
where the $z$-argument of $v_\perp$ is suppressed for simplicity. Integrating both sides with respect to $t$ once and twice, we obtain
\begin{subequations}\label{SCLC}
	\begin{equation}\label{v}
	v_\perp(t) = \frac{e\mathcal{J}}{2m\varepsilon_0} t^2 + \frac{e}{m F } t,
	\end{equation}
	\begin{equation}\label{x}
	x(t) = \frac{e\mathcal{J}}{6m\varepsilon_0}t^3 + \frac{e\mathcal{E}}{2m} t^2.
	\end{equation}
\end{subequations}
The boundary values of Eq. (\ref{SCLC}) are: $x(t_0) = D$, $v_\perp(t_0) = \sqrt{2eV_0/m}$ where $t_0$ is the transit time for emitted electron to travel across the cathode-anode distance of $D$,  $v_\perp(t_0) = \sqrt{2eV_0/m}$ is the final velocity of the emitted electron in reaching the anode and $V_0$ is the bias voltage. Using these boundary values, Eq. (\ref{SCLC}) becomes
\begin{subequations}\label{SCLC2}
	\begin{equation}\label{SCLC2_1}
	V_0 = \frac{e\mathcal{J}}{2m\varepsilon_0} t_0^2 + \frac{e}{m F} t_0,
	\end{equation}
	\begin{equation}\label{SCLC2_2}
	D = \frac{e\mathcal{J}}{6m\varepsilon_0}t_0^3 + \frac{eF}{2m} t_0^2.
	\end{equation}
\end{subequations}
Note that at large $\mathcal{J}$, Eqs. (\ref{SCLC2_1}) and (\ref{SCLC2_2}) can be simultaneously solved to give the CL law,
\begin{equation}\label{CL}
 J_{CL} = \frac{4\varepsilon_0}{9}\sqrt{\frac{2e}{m}}\frac{V_0^{3/2}}{D^2},
\end{equation}
which is a universal relation independent of the emitter material properties.

Equation (\ref{SCLC2_1}) can be solved in a dimensionless form to yield $\bar{t} = \xi \bar{F} / \bar{J}$ where $\xi = -1 + \sqrt{1+ 2\sqrt{2\bar{V}} \bar{F}^2 / \bar{J}}$. The term $\xi$ can be rearranged to give an expression of $\bar{V}$, 
\begin{equation} \label{xi}
\bar{V} = \frac{\bar{F}^4}{8 \bar{J}^2} \left[ \left(\xi + 1\right)^2 - 1 \right]^2 .
\end{equation}
In deriving Eq. $(\ref{xi})$, we have defined the following dimensionless term: $\bar{t} = t_0/\tau_0$, $\bar{F} = F/F_0$, $\bar{\mathcal{J}} = \mathcal{J} / \mathcal{J}_0$, and $\bar{V} = V_0 /U_0$ where the normalization factors are $\tau_0 = \epsilon_0 F_0/\mathcal{J}_0$, $F_0 = 4\sqrt{2m}\Phi_B^{3/2}/3\hbar e$, $D_0 = eF_0\tau_0^2/m$ and $U_0 = F_0 D_0$. The definition of $\mathcal{J}_0$ is dependent on the explicit forms of $\mathcal{J}$ [see Eqs. (\ref{J_dimensionless}) and (\ref{J_norm}) below]. 
Equation (\ref{SCLC2_2}) can be  analytically solved as
\begin{equation}\label{xi_sol}
\xi = h(p) + h(p)^{-1} -1,
\end{equation}
where 
\begin{equation}
h(p) \equiv  \left( \sqrt{ \left(\frac{p}{2}\right)^2 - p } + \frac{p}{2} - 1 \right)^{1/3},
\end{equation}
and $p \equiv 6\bar{D} \mathcal{\bar{J}} / \bar{F}^3$. 
The value of $\xi$ is dependent on the form of the dimensionless FE current densities, $\bar{\mathcal{J}}$. We cast Eqs. (\ref{C_FE}) and (\ref{NC_FE}) into the dimensionless form of $\bar{\mathcal{J}}(\bar{F}) \equiv \mathcal{J}(F) / \mathcal{J}_0$, where
	\begin{equation}\label{J_dimensionless}
	\begin{bmatrix}
	\bar{\mathcal{J}}_{ \text{2DEG} }^{(\text{C})}(\bar{F}) \\
	\bar{\mathcal{J}}_{ \text{Dirac} }^{(\text{C})}(\bar{F}) \\
	\bar{\mathcal{J}}_{ \text{2DEG} }^{(\text{NC})}(\bar{F}) \\
	\bar{\mathcal{J}}_{ \text{Dirac} }^{(\text{NC})}(\bar{F})
	\end{bmatrix}
	= 
	\begin{bmatrix}
	e^ {- \frac{ \bar{\varepsilon}_F }{ \bar{F} } } \\
	e^ {- \frac{ \bar{\varepsilon}_F }{ \bar{F} } } \\
	\lambda \bar{F}\Lambda_{1}\left(\frac{\bar{\varepsilon}_F}{\bar{F}}\right) \\
	\lambda \bar{F}^2 \Lambda_{2}\left(\frac{\bar{\varepsilon}_F}{\bar{F}}\right)
	\end{bmatrix}
	\exp \left( - \frac{ 1}{ \bar{F} } \right),
	\end{equation}
and the normalization factors, $\mathcal{J}_0$'s, are given as,
\begin{equation}\label{J_norm}
\begin{bmatrix}
\mathcal{J}_{0,\text{2DEG} }^{(\text{C})} \\
\mathcal{J}_{0, \text{Dirac} }^{(\text{C})} \\
\mathcal{J}_{0 ,\text{2DEG} }^{(\text{NC})} \\
\mathcal{J}_{0 , \text{Dirac} }^{(\text{NC})}
\end{bmatrix}
=
\frac{g_{s,v}e}{2\pi \hbar^2}
\frac{v(\tilde{\varepsilon}_\perp)}{L_\perp}
\begin{bmatrix}
m^*\varepsilon_F \\
\frac{\varepsilon_F^2}{2v_F^2} \\
\frac{2m^*\Phi_B}{3}  \\
\left( \frac{2\Phi_B}{3 v_F} \right)^2 
\end{bmatrix}.
\end{equation}
Here, $\bar{\varepsilon}_F \equiv 3\varepsilon_F / 2 \Phi_B$, $\Lambda_{1}(x) \equiv 1 - e^{-x} $ and $\Lambda_{2}(x) \equiv x + e^{-x} - 1$. 
In Tables I and II, we summarize the numerical values of $\mathcal{J}_{0}$, $D_0$ and $V_0$ with $\varepsilon_F = 0.075$ eV.

The $\mathcal{\bar{J}}$-$\bar{V}$ characteristics can be solved by coupling $\mathcal{\bar{J}}$ [Eq. (\ref{J_dimensionless})] with $\bar{V}$ [Eq. (\ref{xi})] via the solution of $\xi$ [Eq. (\ref{xi_sol})]. 
Before presenting the general $\mathcal{\bar{J}}$-$\bar{V}$ characteristics, we first consider two asymptotic limits of $\xi \to 0$ and $\xi \to \infty$ where $\xi$ approaches the following relations,
	\begin{subequations}\label{xi_lim}
		\begin{equation}\label{xi_0}
		\lim_{\xi \to 0} \xi = \bar{\mathcal{\bar{J}}}\sqrt{\frac{6\bar{D}}{3\bar{F}^{3}}} ,
		\end{equation}
		\begin{equation}\label{xi_infty}
		\lim_{\xi \to \infty} \xi = \frac{\left( 6\bar{D} \mathcal{\bar{J}}^{2} \right)^{1/3}}{\bar{F}} .
		\end{equation}
	\end{subequations}
	Correspondingly, $\bar{V}$ in Eq. (\ref{xi}) becomes
	\begin{subequations}\label{V_lim}
		\begin{equation}\label{V_0}
		\bar{V}\left(\xi \to 0\right) \approx \frac{\bar{F}^4  }{2 \bar{\mathcal{J}}^2 } \xi^2,
		\end{equation}
		\begin{equation}\label{V_infty}
		\bar{V}(\xi \to \infty) \approx \frac{\bar{F}^4 } {8\bar{\mathcal{J}}^2} \xi^4.
		\end{equation}
	\end{subequations}
	By combining Eqs. (\ref{xi_0}) and (\ref{V_0}), we obtain $\bar{F} = \bar{V}/\bar{D}$ at the $\xi \to 0$ limit, i.e. the $\bar{\mathcal{J}}$-$\bar{V}$ characteristics can be obtained straightforwardly from Eq. (\ref{J_dimensionless}) by substituting $\bar{F}\to\bar{V}/\bar{D}$. This simply recovers the FE $\bar{\mathcal{J}}$-$\bar{V}$ characteristics as,
	\begin{equation}\label{xi_0_J}
	\lim_{\xi \to 0}
	\begin{bmatrix}
	\bar{\mathcal{J}}_{ \star \text{2DEG} }^{(\text{C})}(\bar{V}) \\
	\bar{\mathcal{J}}_{ \star \text{Dirac} }^{(\text{C})}(\bar{V}) \\
	\bar{\mathcal{J}}_{ \star \text{2DEG} }^{(\text{NC})}(\bar{V}) \\
	\bar{\mathcal{J}}_{ \star \text{Dirac} }^{(\text{NC})}(\bar{V})
	\end{bmatrix}
	= 
	\begin{bmatrix}
	e^ {- \frac{ \bar{\varepsilon}_F \bar{D} }{ \bar{V} } } \\
	e^ {- \frac{ \bar{\varepsilon}_F \bar{D} }{ \bar{V} } } \\
	\lambda \frac{\bar{V}} {\bar{D} }\Lambda_{1}\left( \frac{\varepsilon_F\bar{D}}{\bar{V}} \right) \\
	\lambda \frac{\bar{V}^2} {\bar{D}^2 } \Lambda_{2}\left( \frac{\varepsilon_F\bar{D}}{\bar{V}} \right) 
	\end{bmatrix}
	\exp \left( - \frac{ \bar{D} }{ \bar{V} } \right).
	\end{equation}
	In the limit of $\xi \to \infty$, the combination of Eqs. (\ref{xi_infty}) and (\ref{V_infty}) yields
	\begin{equation}\label{xi_infty_J}
	\lim_{\xi \to \infty}
	\mathcal{\bar{J}}_{\text{CL}}(\bar{V}) = \frac{2\sqrt{2}}{3} \frac{\bar{V}^{3/2} } { \bar{D}^2},
	\end{equation}
	which is the dimensionless form of the CL law [see Eq. (\ref{CL})].

\begin{figure*}
	\includegraphics[scale = 0.68]{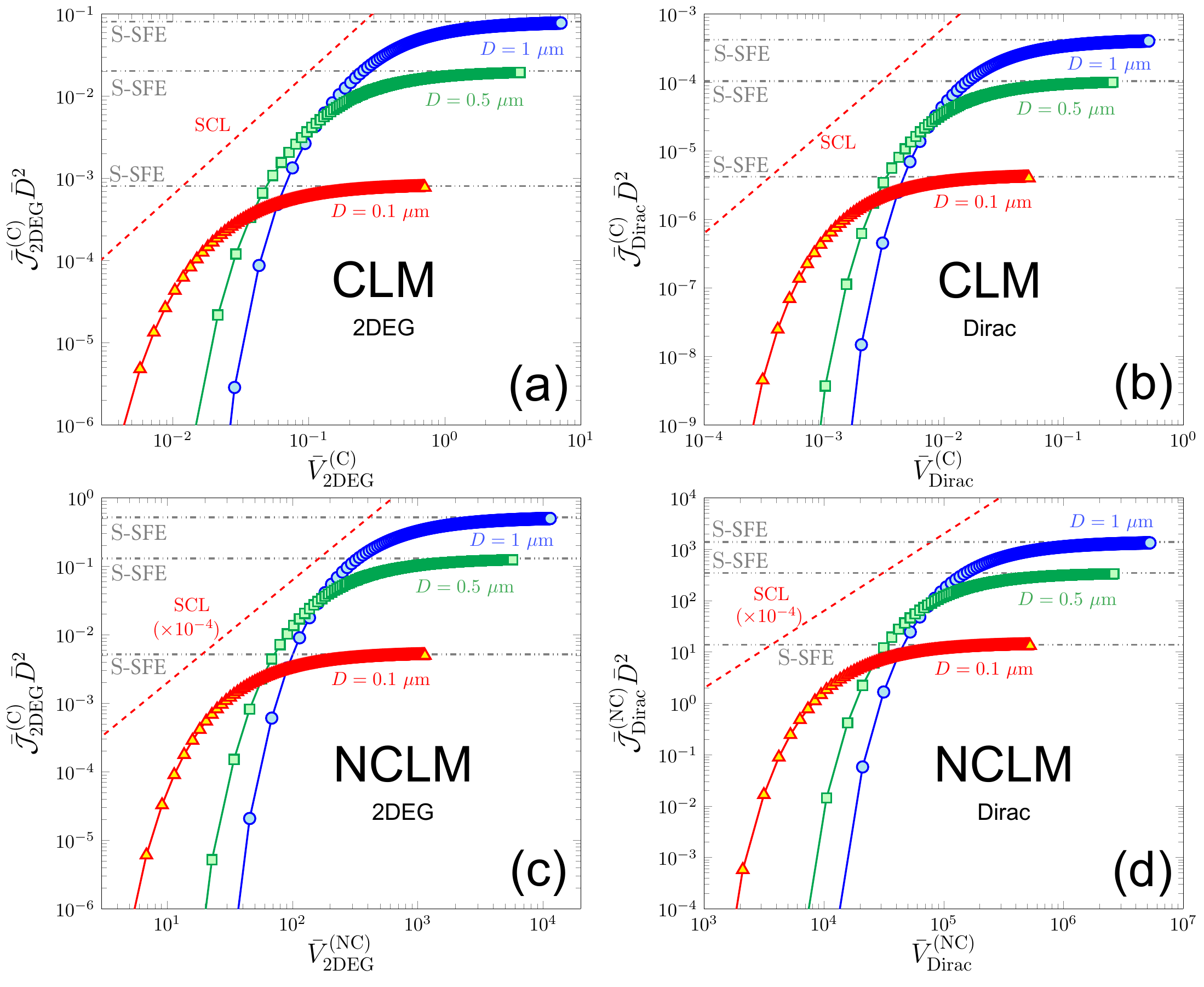}
	\caption{Dimensionless $\bar{\mathcal{J}}$-$\bar{V}$ characteristics in the presence of space-charge. (a) and (b) show the CLM current-voltage characteristics of parabolic and linear lateral energy dispersion, respectively. (c) and (d) show the NCLM counterpart of (a) and (b), respectively. The red dashed line represents the universal SCL limit and the dash-dotted lines represents the saturated emission limit (labeled as `S-SFE'), i.e. $\bar{\mathcal{J}}_{\text{Dirac}}^{(\text{NC})} \bar{D}^2 = \bar{\varepsilon}_F^2 \bar{D}^2/2$ and $\bar{\mathcal{J}}_{\text{2DEG}}^{(\text{NC})} \bar{D}^2 = \bar{\varepsilon}_F \bar{D}^2$. The Fermi level is set to $\varepsilon_F = 0.075$ eV. Note that the SCL limits in (c) and (d) are down-shifted by a factor of $10^{-4}$.}
\end{figure*}

\begin{figure*}
	\includegraphics[scale=1.1]{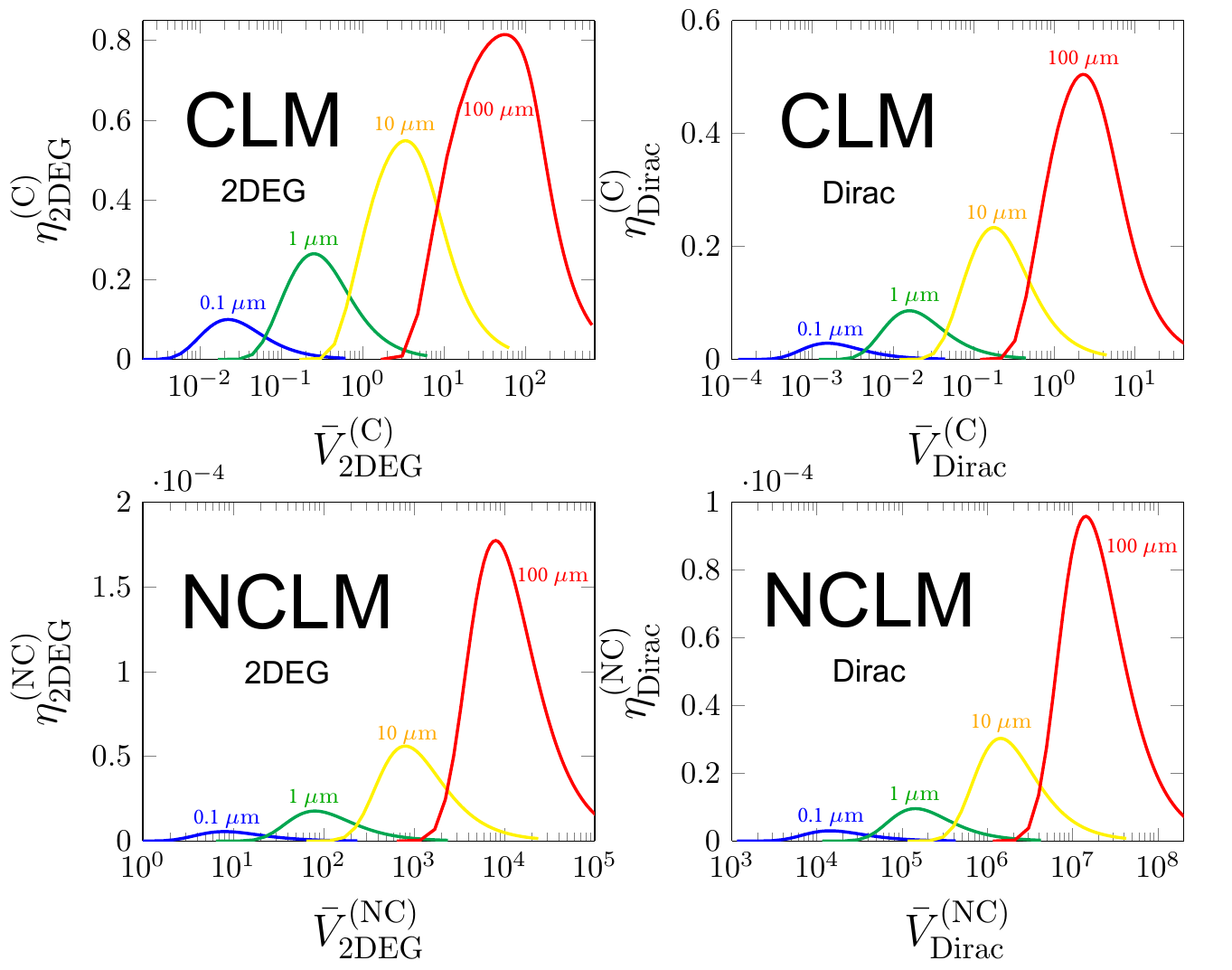}
	\caption{Absence of space-charge-limited (SCL) emission in 2D material surface field emission. The $\eta$'s are plotted as a function of $\bar{V}$ at $D = [0.1, 1, 10, 100]$ $\mu$m with $\varepsilon_F = 0.075$ eV. (a) CLM model with parabolic-$\varepsilon_\parallel$; (b) CLM model with linear-$\varepsilon_\parallel$. (c) and (d): same as (a) and (b) for NCLM models. In all cases, $\eta<1$ suggests that full SCL emission is absent in 2D material field emitter. }
\end{figure*}

In Figs. 7, the $\mathcal{\bar{J}}$-$\bar{V}$ curves of CLM [Figs. 7(a) and (b)] and NCLM [Figs. 7(c) and (d)] models are plotted by simultaneously solving Eqs. (\ref{xi}), (\ref{xi_sol}) and (\ref{J_dimensionless}).
Three vacuum gap distances are used: 0.1 $\mu$m, 0.5 $\mu$m and 1 $\mu$m. The Fermi level is set to $\varepsilon_F = 0.075$ eV. 
At small $\bar{V}$, the $\mathcal{\bar{J}}^{(\text{C})}$ increases approximately exponentially with $\bar{V}$ as a result of the field-induced quantum tunneling. 
The $\mathcal{\bar{J}}^{(\text{C})}$-$\bar{V}$ curves shift towards the SCL limit (dashed curve in Fig. 7) as $\bar{V}$ increases due to the building up of space charge in the vacuum gap.
However, when $\bar{V}$ is further increased, the $\mathcal{\bar{J}}^{(\text{C})}$ curves reach a plateau without dwelling into the full SCL regime.
This peculiar absence of full SCL flow can be explained by the saturated emission, which is a source-limited mechanism. 
At sufficiently high voltage, the emission current densities saturated. In this case, the 2D-material-based cathode fails to supply the required magnitude of charge current to establish a full SCL flow. This leads to a remarkable consequence: \emph{SCL emission is completely absent} in 2D-material-based vacuum diode at nanometer and micrometer scales, with practical bias voltage range of $V <1$ kV. 

	\begin{widetext}
		\begin{center}
			\begin{table}[t]
				\caption{Numerical values of the normalization constant for CLM model evaluated at $\Phi_{B0} = 4.5$ eV and $\varepsilon_F = 0.075$ eV. }
				\begin{tabular}{l|l|l|l|l|l|l}
					\hline \hline
					Normalization constant & $\mathcal{J}_{0,\text{2DEG}}^{(\text{C})}$ $(\text{A/m}^2)$     & $\mathcal{J}_{0,\text{Dirac}}^{(\text{C})}$ $(\text{A/m}^2)$  & $D_{0,\text{2DEG}}^{(\text{C})}$ (m) & $D_{0,\text{Dirac}}^{(\text{C})}$ (m) & $U_{0,\text{2DEG}}^{(\text{C})}$ (V) & $U_{0,\text{Dirac}}^{(\text{C})}$ (m) \\ \hline
					Numerical value    & $1.65\times 10^{13}$ &  $8.86\times 10^{12}$    &    $1.30\times 10^{-5}$   & $4.52\times10^{-5}$ & 8.28 $\times 10^5$ & $2.87 \times 10^{6}$ \\ \hline\hline
				\end{tabular}
			\end{table}
		\end{center}
		\begin{center}
			\begin{table}[t]
				\caption{Numerical values of the normalization constant for NCLM model evaluated at $\Phi_{B0} = 4.5$ eV and $\varepsilon_F = 0.075$ eV.}
				\begin{tabular}{l|l|l|l|l|l|l}
					\hline \hline
					Normalization constant& $\mathcal{J}_{0,\text{2DEG}}^{(\text{NC})}$ $(\text{A/m}^2)$     & $\mathcal{J}_{0,\text{Dirac}}^{(\text{NC})}$ $(\text{A/m}^2)$  & $D_{0,\text{2DEG}}^{(\text{NC})}$ (m) & $D_{0,\text{Dirac}}^{(\text{NC})}$ (m) & $U_{0,\text{2DEG}}^{(\text{NC})}$ (V) & $U_{0,\text{Dirac}}^{(\text{NC})}$ (m) \\ \hline
					Numerical value    & $6.49\times 10^{14}$ &  $2.74\times 10^{16}$    &    $8.42\times 10^{-9}$   & $4.72\times10^{-12}$ & 535 & 0.30 \\ \hline\hline
				\end{tabular}
			\end{table}
		\end{center}
	\end{widetext}

The strength of SCL effect can be quantified by the ratio, $\eta \equiv \bar{\mathcal{J}}(\bar{V})/\bar{\mathcal{J}}_{\text{CL}}(\bar{V})$, where $\eta = 1$ signifies the onset of full SCL flow. In Fig. 8, the $\eta$'s of CLM and NCLM models are plotted for ten values of $D$ ranging from $0.1$ $\mu$m to 100 $\mu$m. The $\eta$ curves briefly peak and decline due to the onset of saturated emission. For CLM models, the maximum $\eta$ with $D = 1$ $\mu$m are found to be $\eta_{\text{2DEG}}^{(\text{C})} \approx 0.26$ and $\eta_{\text{Dirac}}^{(\text{C})} \approx 0.085$ [Figs. 8(a) and (b)], thus suggesting only a maximal of $26$\% and $8.5$\% of the SCL limit is achievable in 2D material with parabolic and linear lateral energy dispersion, respectively. This suppression is even more dramatic for NCLM models [Figs. 8(c) and (d)] in which the maximum $\eta$ are in the order of $\eta^{(\text{NC})}\sim 10^{-5}$ for both types of lateral energy dispersions. This negligibly small magnitude of $\eta \ll 1$ suggests that SCL flow is practically absent if the SFE process does not conserve lateral momentum.
Even for a vacuum gap size of 100 $\mu$m which requires an exceedingly large bias voltage of $V > 10^6$ V, the CLM model can maximally achieve $\eta_{\text{2DEG}}^{(\text{C})} \approx 0.80$ and $\eta_{\text{Dirac}}^{(\text{C})} \approx 0.50$. The onset of full SCL emission remains out of reach.
In calculating $\eta^{(\text{NC})}$, we have estimated $\lambda = 10^{-4}$. By tuning to the NCLM process to the unrealistic full-strength value of $\lambda = 1$, we have $\eta^{(\text{NC})}_{\text{2DEG}} \approx 0.70 $ and $\eta^{(\text{NC})}_{\text{Dirac}} \approx 0.50$, which are still fractions of the full SCL limit.

We now estimate the value of bias voltage at which the saturated emission current density can be observed experimentally. The onset of the saturated emission can be estimated from Fig. 8 as the bias voltage at which the $\eta$ curve peaks. For CLM models with $D = 0.1$ $\mu$m, the onset of saturated emission occurs at $\bar{V}_{\text{2DEG}}^{(\text{C})} \gtrsim 0.02$ and $\bar{V}_{\text{Dirac}}^{(\text{C})} \gtrsim 0.002$ [Figs. 8(a) and (b)]. Consider a conservative value of field enhancement factor of $b_F = 20$, we obtain $V_{\text{2DEG}}^{(\text{C})}\gtrsim 830$ V and $V_{\text{Dirac}}^{(\text{C})}\gtrsim 280$ V in SI unit. For NCLM models, the onset of saturated emission occurs at $\bar{V}_{\text{2DEG}}^{(\text{NC})} \gtrsim 10$ and $\bar{V}_{\text{Dirac}}^{(\text{NC})} \gtrsim 10^4$ [Figs. 8(c) and (d)], which correspond to $V_{\text{2DEG}}^{(\text{NC})} \gtrsim 270$ V and $V_{\text{Dirac}}^{(\text{NC})}\gtrsim 150$ V. These values are well within the achievable voltage range in experiments. Thus, the saturated emission and the absence of SCL flow shall be observable in monolayer and bilayer graphene surface field emitters. 

\section{Discussion}

\subsection{Physical interpretation of the `time'-like parameter in Sinha-Lee's `time'-constant-based thermionic emission model}

We now discuss the relation between the generalized formalism for 2D material electron emission developed above and Sinha-Lee's `time'-constant-based TE model for graphene \cite{sinha}.
We shall show that our model can recover the physical meaning of the arbitrarily-defined `time'-parameter, namely $\tau$, in Sinha-Lee's model.

In Sinha-Lee's model, the $\varepsilon_\perp$-integral in the emission current density equation is replaced by the `time'-like parameter, $\tau$, i.e.
\begin{eqnarray}\label{SL}
\mathscr{J}_{\text{SH}} &=& \frac{e}{\tau} \int D_{\text{Dirac}}(\varepsilon_\parallel) \Theta(\varepsilon_\parallel - \Phi_B) f(\varepsilon_\parallel) d\varepsilon_\parallel \nonumber \\
&=& \frac{1}{\tau} \frac{g_{s,v}e}{2\pi \hbar^2 v_F^2 } \left(k_BT\right)^2 \left[1 + \frac{\Phi_B}{k_BT}\right] \exp\left( - \frac{\Phi_B - \varepsilon_F}{k_BT} \right) \nonumber \\
\end{eqnarray}
In obtaining Eq. (\ref{SL}), the over-barrier transmission probability is assumed to take the form of $\mathcal{T}(\varepsilon_\parallel) = \Theta(\varepsilon_\parallel -\Phi_B)$. This $\varepsilon_\parallel$-dependent transmission probability, rather than $\varepsilon_\perp$-dependent one, immediately suggests that the Sinha-Lee TE model belongs to the class of NCLM model.
The physical interpretation of $\tau$ remains fuzzy in this model. 
As $\tau$ carries the physical dimension of time, it is hypothesized that it represents the time scale of how fast electrons are extracted from graphene surface.

Here, based on the NCLM model developed in Eq. (\ref{J_gen_2D_NC}), a concrete physical meaning of $\tau$ can be recovered.
By using the standard TE approximations of $\mathcal{T}(\varepsilon_\parallel) \approx  \Theta(\varepsilon_\parallel - \Phi_B)$ and $f(\varepsilon_\parallel) \approx \exp\left[ \left(\varepsilon_\parallel - \varepsilon_F\right)/k_BT \right]$, the NCLM model of graphene TE can be solved as
\begin{eqnarray}\label{NC_TE}
\mathscr{J}^{(\text{NC})}_{\text{Dirac}} &=& \lambda \frac{ e v(\tilde{\varepsilon}_\perp)}{L_\perp} \int_{\Phi_B}^{\infty} D_{\text{Dirac}}(\varepsilon_\parallel) \exp\left( - \frac{\varepsilon_\parallel - \varepsilon_F}{k_BT}\right) d\varepsilon_\parallel \nonumber \\
&=&  \lambda \frac{v(\tilde{\varepsilon}_\perp)}{L_\perp} \frac{ g_{s,v} e }{2\pi  \hbar^2v_F^2} \left(k_BT\right)^2\left[ 1 + \frac{\Phi_B}{k_BT} \right]  \nonumber \\
&& \times \exp\left( - \frac{\varepsilon_\parallel - \varepsilon_F}{k_BT}\right)
\end{eqnarray}
By equating Eqs. (\ref{SL}) and (\ref{NC_TE}), we can write the `time'-like parameter in terms of physical terms, i.e.
\begin{equation}\label{tau}
\tau = \frac{L_\perp}{\lambda v(\tilde{\varepsilon}_\perp)}
\end{equation}
Here, the physics of $\tau$ becomes apparent: $\tau$ is directly related to the strength of NCLM process, the 2D material thickness and the out-of-plane bound state electron velocity. This suggests that interpreting $\tau$ as the time scale of how fast electrons are extracted from the thermionic emitter is not appropriate since $L_\perp$ is the 2D material thickness rather than the separation between the electron emitter and the collector. Using $L_\perp = 0.335$ nm, $\tilde{\varepsilon}_\perp = 38.3$ eV, $v(\tilde{\varepsilon}_\perp) = \sqrt{2\tilde{\varepsilon}_\perp/m} = 3.67 \times 10^6$ m/s and setting the NCLM process to the full strength of $\lambda = 1$, the \emph{lower limit} of $\tau$ can be predicted as $\tau_{\text{min}} \approx 0.091 $ fs.

The value of $\tau$ varies depending on the quality of the interface. For graphene/silicon and graphene/Pd contacts, the experimentally extracted $\tau$'s are $46.2$ ps and $0.13$ ps, respectively \cite{sinha}. These values correspond to $\lambda \approx 1.87 \times 10^{-6}$ and $\lambda \approx 6.64 \times 10^{-4}$, respectively. In the above sections, we choose $\lambda = 10^{-4}$ in the calculation of NCLM emission models. Note that even in the case of unrealistic $\lambda = 1$, the unconventional saturated emission and the absence of full SCL flow reported in this work remain robust. Such effects becomes even more profound for smaller value of $\lambda < 10^{-4}$.

\subsection{Summary and outlooks}

For the ease-of-use of the models developed above in the analysis of experimental data, we summarize in Table III the field-induced surface electron emission models in \emph{empirical forms} where only the $F$-dependences are explicitly shown. 
We note that the model can be further improved, for example, by using the exact expression of the triangular barrier tunneling probability, including the image potential and generalization of the model to include thermionic contributions \cite{forbes2, forbes3, jensen, jensen2}. 
Nonetheless, we expect these refinements to only quantitatively alter the results. 
The dimensionality-induced saturated emission and the strong suppression of SCL electron flow shall qualitatively remains the same. 
Finally, we remark that although the two versions of vertical emission models developed above, i.e. with and without lateral momentum conservation, exhibit different forms in general, they both converge to the same unconventional scaling relation of $\log(J) \propto 1/F$ at high field. This suggests that the modified FN-plot in Eq. (\ref{CLM_FN_plot}) shall be universally applicable to the cases of both conversing and non-conserving lateral momentum. Further experimental and \emph{ab initio} computational works are required to pin down the mechanism (i.e. CLM or NCLM) underlying the field-induced vertical electron emission from the surface of 2D materials. 

\begin{widetext}
	\begin{center}
		\begin{table}
			\caption{CLM and NCLM models of surface electron field emission from 2D materials in the empirical forms. The terms $\mathcal{A}$, $\mathcal{B}$ and $\mathcal{C}$ are independent of electric field strength, $F$. }
			\begin{tabular}{l | l | l | l }
				\hline\hline
				Model  & Lateral energy dispersion ($\varepsilon_\parallel$) & Empirical form     & Full expression  \\ \hline \hline 
				FN Law & Parabolic & $\mathcal{J}_{\text{FN}} = \mathcal{C} F^2 \exp \left( -\frac{\mathcal{B}}{F} \right)$ & Eq. (1b) \\ \hline
				CLM & parabolic & $\mathcal{J}_{\text{2DEG}}^{(\text{C})} = \mathcal{C} \exp \left( -\frac{\mathcal{B}}{F} \right)$ & Eq. (14a)  \\ 
				& linear & $\mathcal{J}_{\text{Dirac}}^{(\text{C})} = \mathcal{C} \exp \left( -\frac{\mathcal{B}}{F} \right)$ &  Eq. (14b)       \\	\hline
				NCLM & Parabolic & $\mathcal{J}_{\text{2DEG}}^{(\text{NC})} = \mathcal{C}F \left[ 1 - \exp\left( -\frac{\mathcal{A}}{F} \right) \right] \exp \left( -\frac{\mathcal{B}}{F} \right)$ & Eq. (17a)  \\
				& Linear &   $\mathcal{J}_{\text{Dirac}}^{(\text{NC})} = \mathcal{C}F^2 \left[ \frac{\mathcal{A}}{F} - 1 + \exp\left( -\frac{\mathcal{A}}{F} \right) \right] \exp \left( -\frac{\mathcal{B}}{F} \right)$ & Eq. (17b) \\
				& & & \\
				(Low-field regime) & Parabolic & $\mathcal{J}_{\text{2DEG}}^{(\text{NC})} = \mathcal{C} F \exp \left( -\frac{\mathcal{B}}{F} \right)$ & Eq. (18a) \\
				& Linear & $\mathcal{J}_{\text{Dirac}}^{(\text{NC})} = \mathcal{C} F \exp \left( -\frac{\mathcal{B}}{F} \right)$ &  Eq. (18b) \\
				& & & \\
				(High-field regime) & Parabolic & $\mathcal{J}_{\text{2DEG}}^{(\text{NC})} = \mathcal{C} \exp \left( -\frac{\mathcal{B}}{F} \right)$ & Eq. (20a) \\
				& Linear & $\mathcal{J}_{\text{Dirac}}^{(\text{NC})} = \mathcal{C} \exp \left( -\frac{\mathcal{B}}{F} \right)$ &  Eq. (20b) \\ \hline\hline
			\end{tabular}
		\end{table}
	\end{center}
\end{widetext}

\section{Conclusion}

In conclusion, we have developed analytical and semi-analytical models of field-induced vertical electron emission from the surface of 2D materials by explicitly taking into account the reduced dimensionality, non-parabolic energy dispersion, non-conservation of the lateral momentum, finite-temperature and space-charge-limited effects. 
Our proposed models are more consistent with the physical properties of 2D-material-based field emitter, which are emerging rapidly in the past decade.
We show that the conventional Fowler-Nordheim paradigm for 3D-bulk-material-based field emitter is no longer valid for 2D materials and it should be replaced by the modified models developed above. 
The reduced-dimensionality of 2D material leads to the occurrence of saturated emission and the strong suppression of space-charge-limited electron flow in nanometer-and micrometer-scale vacuum devices.
The generalized 2D material field emission models can be readily applied to describe the charge transport physics in 2D-material-based electrical contact, which is crucially important in the design and engineering of 2D-material-based nanoelectronic and optoelectronic devices, and will be discussed in a separate work.

\section*{acknowledgment}
	
We thank Ji Xu and Qilong Wang for helpful discussions and for providing the experimental data of graphene surface field emission. We thank Shi-Jun Liang for stimulating discussions. This work is supported by A*STAR-IRG (A1783c0011) and AFOSR AOARD (FA2386-17-1-4020).


\begin{thebibliography}{99}
\bibitem{emission_const} The emission constants are given as $A_{\text{RD}} = 4\pi me/h^3$, $A_{\text{FN}} = e^2/8\pi h\Phi_B$ and $B_{\text{FN}} = 4\sqrt{2m}\Phi_B^{3/2}/3\hbar e$ where $h$ and $\hbar$ are the Planck constant and reduced Planck constant, respectively.
\bibitem{foot_1} The Sommerfeld expansion is given as
	%
	\begin{eqnarray}
	\int_0^{\infty} \mathcal{H}(\varepsilon_\parallel) f(\varepsilon_\parallel) d\varepsilon_\parallel &\approx& \int_0^{\varepsilon_F} \mathcal{H}(\varepsilon_\parallel) d\varepsilon_\parallel + \left( \mu - \varepsilon_F \right) \mathcal{H}(\varepsilon_F) \nonumber \\
	&& + \frac{\left(\pi k_BT\right)^2}{6} \left.  \frac{d \mathcal{H}(\varepsilon_\parallel)}{d\varepsilon_\parallel}\right|_{\varepsilon_\parallel = \varepsilon_F}, \nonumber
	\end{eqnarray}
	%
	where $\mathcal{H}(\varepsilon_\parallel)$ is a given function of $\varepsilon_\parallel$ and $\mu$ is the chemical potential which can be approximated, up to second order in $T$, from the Sommerfeld expansion by setting $\mathcal{H}(\varepsilon_\parallel) = D(\varepsilon_\parallel)$. For parabolic and linear energy dispersion, $\mu$ is obtained as $\mu_{\text{2DEG}} = \varepsilon_F$ and $\mu_{\text{Dirac}} = \varepsilon_F - (\pi k_BT)^2/6\varepsilon_F$, respectively.
\end{thebibliography}
\end{document}